\documentclass[aps,preprint,floatfix]{revtex4}

\usepackage{dcolumn}
\usepackage{bm}
\usepackage{graphicx}
\graphicspath{{Images/}}
\usepackage{epstopdf}

\begin{document}

\title{Hyperons: the strange ingredients of the nuclear equation of state}

\author{Isaac Vida\~na}
\affiliation{Istituto Nazionale di Fisica Nucleare, Dipartimento di Fisica, Universit\`a di Catania, Via Santa Sofia 64, I-95123, Catania, Italy}

\begin{abstract}

In this article we will review the role and properties of hyperons in finite and infinite nuclear systems.
In particular, we will revise different production mechanisms of hypernuclei, as well as several aspects of hypernuclear $\gamma$-ray spectroscopy, and the weak decay modes of hypernuclei. Then we will discuss the construction of hyperon-nucleon and hyperon-hyperon interactions on the basis of the meson-exchange
and chiral effective field theories. Recent developments based on the so-called V$_{low\,\, k}$ approach and lattice QCD will also be adressed. Finally, we will go over some of the effects of hyperons on the properties of neutron and proto-neutron stars with an emphasis on the so-called "hyperon puzzle'', {\it i.e.,} the problem of the strong softening of the equation of state, and the consequent reduction of the maximum mass, induced by the presence of hyperons, a problem which has become more intringuing and difficult to solve due the recent measurements of $\sim 2M_\odot$ millisecond pulsars. We will discuss some of the solutions proposed to tackle this problem. We will also re-examine the role of hyperons on the cooling properties of newly born neutron stars and on the development of the so-called r-mode instability.

\end{abstract}

\maketitle



\section{Introduction}

The presence of hyperons ({\it i.e.} baryons with strange content) in finite and infinite nuclear
systems constitutes a unique prove of the deep nuclear interior which gives us the oportunity to study  
baryon-baryon interactions from an enlarge perspective and to extend, in this way, our present knowledge of conventional nuclear physics to the SU(3)-flavor sector \cite{lenske18}. One of the goals of hypernuclear physics \cite{gal16}  is precisely to relate hypernuclear observables with the underlying bare hyperon-nucleon (YN) and hyperon-hyperon (YY) interactions. Nevertheless, contrary to the nucleon-nucleon (NN) interaction, which is fairly well known due to the large number of existing scattering data and measured properties of nuclei, YN and YY interactions are still poorly constrained. The experimental difficulties associated with the short lifetime of hyperons and the low intensity beam fluxes have limited the number of $\Lambda$N
and $\Sigma$N events to several hundred \cite{ln1,ln2,ln3,ln4,ln5} and that of $\Xi$N to very few. In the case of the YY interaction the situation is even worse because no scattering data exists at all. Although this limited amount of data is not enough to fully constrain the YN and YY interactions, complementary information on them can be obtained from the study of hypernuclei, bound systems composed of neutron, protons and one or more hyperons. Hypernuclei were discovered by Danysz and Pniewski \cite{dan53} in 1952 with the observation of a hyperfragment in a balloon-flown emulsion stack. Since then the use of high-energy accelerators as well as modern electronic counters have allowed the identification of 
more than 40 single $\Lambda$-hypernuclei, and few double $\Lambda$ \cite{dl,dl1,dl2,dl3,dl4,dl5,dl6,dl7}
and single-$\Xi$ \cite{kau,naka} ones have been identified. On the contrary, the existence of single $\Sigma$-hypernuclei has not been experimentally confirmed yet without ambiguity (see {\it e.g.,} Refs.\
\cite{bertini80,bertini84,bertini85,piekarz82,yamazaki85,tang88,bart99,hayano89,nagae98}
) suggesting that the $\Sigma$-nucleon interaction is most probably repulsive \cite{dgm89,batty1,batty2,batty3,mares95,dabrowski99,noumi02,saha04,harada05,harada06}.

Attempts to derive the hyperon properties in a finite nucleus have followed several approaches. Traditionally, hypernuclei have been reasonably well described by a shell-model picture using effective $\Lambda$-nucleus mean field potentials of the Woods--Saxon type that reproduce quite well the measured hypernuclear states of medium to heavy hypernuclei \cite{ws1,ws1b,ws2,ws3}. Non-localities and density dependent effects, included in non-relativistic Hartree--Fock calculations using Skyrme-like YN interactions \cite{skhf1,skhf2,skhf3,skhf4,skhf4b,skhf5,skhf5b,skhf5c,skhf5d} improve the overall fit to the single-
particle binding energies. The properties of hypernuclei have also been studied in a relativistic framework, such as Dirac phenomenology, where the hyperon-nucleus potential is derived from the nucleon-nucleus one \cite{dirac1,dirac2}, or relativistic mean field theory \cite{rmf1,rmf2,rmf3,rmf4,rmf5,rmf6,rmf7,rmf8,rmf8b,rmf9}. Microscopic hypernuclear structure calculations, which can provide the desired link between the hypernuclear observables and the bare YN interaction, are also available. They are based on the construction of an effective YN interaction (G-matrix) which is obtained from the bare YN one by solving the Bethe--Goldstone equation. In earlier microscopic calculations, Gaussian parametrizations of the G-matrix calculated in nuclear matter at an average density were employed \cite{g1,g1b,g1c,g1d}. A G-matrix calculated in finite nuclei was used to study the single-
particle energy levels in various hypernuclei \cite{g2}. Nuclear matter G-matrix elements were also used as an effective interaction in a calculation of the $^{17}_\Lambda$O spectrum \cite{g3}. The s- and p-wave $\Lambda$ single-particle properties for a variety of $\Lambda$-hypernuclei from $^5_\Lambda$He to $^{208}_\Lambda$Pb where derived in Refs.\ \cite{morten96,vidana98,vidana00} by constructing a finite nucleus YN G-matrix from a nuclear matter G-matrix. Recently, a Quantum Monte Carlo calculation of single- and double-$\Lambda$ hypernuclei has also been done using two- and three-body forces between the $\Lambda$ and the nucleons \cite{lonardoni1,qmchyp}. In most of these approaches, the quality of the description of hypernuclei relies in the validity of the mean field picture. However, the correlations induced by the YN interaction can substantially change this picture and, therefore, should not be ignored. Very recently, the author
of the present review has studied the spectral function of the $\Lambda$ hyperon in finite nuclei \cite{vidana17}, showing that the $\Lambda$ is less correlated than the nucleons in agreement with the  idea that it maintains its identity inside the nucleus. The results of this study show also that in hypernuclear production reactions the $\Lambda$ hyperon is formed mostly in a quasi-free state.

Despite hypernuclear matter is an idealized physical system, its study has also attracted the attention of many authors in connection with the physics of neutron star interiors \cite{shapiro,shapiro2,shapiro3}. The interior of neutron stars is dense enough to allow for the appearance of new particles with strangeness content besides the conventional nucleons and leptons by virtue of the weak equilibrium. There is a growing evidence that hyperons appear as the first strange baryons in neutron star at around twice normal nuclear saturation density. Properties of neutron stars are closely related to the underlying Equation of State (EoS) of matter at high densities. The theoretical determination of the  hypernuclear matter EoS is therefore an essential step towards the understanding of these properties which
can be affected by the presense of strangeness. Conversely, the comparison of the theoretical predictions 
for these properties with astrophysical observations can provide strong constraints on the YN and YY interactions. Therefore, a detailed knowledge of the EoS of hypernuclear matter over a wide range of densities is requiered. This is a very hard task from the theoretical point of view. Traditionally, two types of approaches have been followed to describe the baryon-baryon interaction in the nuclear medium and, to construct from it the nuclear EoS: phenomenological and  microscopic approaches.

Phenomenological approaches, either relativistic or non-relativistic, are based on effective density-dependent interactions which typically contain a certain number of parameters adjusted to reproduce nuclear and hypernuclear observables, and neutron star properties. Skyrme-type interactions \cite{skyrme} and relativistic mean field (RMF) models \cite{serot} are among the most commonly used ones within this type of approaches. Skyrme-type forces have gained so much importance because they reproduce the binding energies and the nuclear radii over the whole periodic table with a reasonable set of parameters. Balberg and Gal \cite{shf1,shf2} derived an analytic effective EoS using density-dependent baryon-baryon potentials based on Skyrme-type forces incluing hyperonic degrees of freedom. The features of this EoS rely on the properties of nuclei for the NN intertaction, and mainly on the experimental data from hypernuclei for the YN and YY ones. It reproduces typical properties of high-density matter found in theoretical microscopic models. RMF models treat the baryonic and mesonic degrees of freedom explicitely. They are fully relativistic and are, in general, easier to handle because they only involve local densities and fields. The EoS of dense matter with hyperons was first described within the RMF by Glendenning \cite{glend,glend2,glend3,glend4} and then by many other authors (see {\it e.g.} Ref. \cite{rmfa,rmfb,rmfc,rmfd}). The parameters of this approach are fixed by the properties of nuclei and nuclear bulk matter for the nucleonic sector, whereas the couping constants of the hyperons are fixed by symmetry relations and hypernuclear observables. 

Microscopic approaches, on the other hand, are based on realistic two-body baryon-baryon
 interactions that describe the scattering data in free space. These realistic 
interactions have been mainly constructed within the framework of a meson-exchange  theory \cite{nagels73, machleidt87,na78,nij1,nij2,nijmegenb,nij3,nij4, juelich89,juelich04}, although in the last years a new approach based on chiral perturbation theory has emerged as a powerful tool
\cite{wei91,wei91b,entem03,epelbaum05,xeft1,xeft2,xeft3,polinder06,haidenbauer13}. In order to obtain the EoS one has to solve then the very complicated many-body problem \cite{mbp}. A great difficulty of this problem lies in the treatment of the repulsive core, which dominates the short-range behavior of the interaction. Although different microscopic many-body methods have been extensively used to the study of nuclear matter, up to our knowledge, only the Brueckner--Hartree--Fock (BHF) approximation \cite{micro,micro2,micro3,micro4,micro5,micro6,micro7,micro8} of the Brueckner--Bethe--Goldstone theory, the $V_{low \> k}$ approach \cite{vlowk}, the Dirac--Brueckner--Hartree--Fock theory \cite{dbhf1,dbhf2}, and very recently the Auxiliary Field Diffusion Monte Carlo method \cite{qmc}, have been extended to the hyperonic sector.

To finish this introduction, we would like to stress that although hypernuclear physics is almost sixty years old, it is still a very active field of research. New experimental facilitires under construction at FAIR/GSI, JLAB, J-JARC and other sites will soon allow for a more precise determination of the properties of hyperon-nucleon and hyperon-hyperon forces than is currently availble. In addition, the recent detection by the LIGO/Virgo collaborations of the gravitational waves emitted in the coalescence of two neutron stars \cite{ligo} opens a new era of astrophysical observation that in the near future will allow to constrain further the dense matter EoS.    

The manuscript is organized in the following way. Production mechanisms of single- and double-$\Lambda$ hypernuclei, several aspects of $\gamma$-ray hypernuclear spectroscopy and weak decay modes of hypernuclei are discussed in Sec.\ \ref{sec:p}. In Sec.\ \label{sec:models} we present different approaches to derive the hyperon-nucleon and hyperon-hyperon interactions based on the meson-exchange and chiral effective field theories. Recent developments from so-called V$_{low\,\,k}$ approach and lattice QCD are also reviewed in this section. The influence of hyperons on the properties of neutron stars is revised in Sec.\ \ref{sec:stars}. Finally, a summary is presented in Sec.\ \ref{sec:conclusions}.


\section{Production, Spectroscopy and Weak Decay of Hypernuclei}
\label{sec:p}

In the following we will briefly describe different production mechanisms of hypernuclei. After that we will discuss some aspects of hypernuclear $\gamma$-ray spectroscopy, and we will finish this section by presenting the different weak decay modes of hypernuclei.


\subsection{Production mechanisms of hypernuclei}

\begin{figure}[t!]
\begin{center}
\includegraphics[width=12cm]{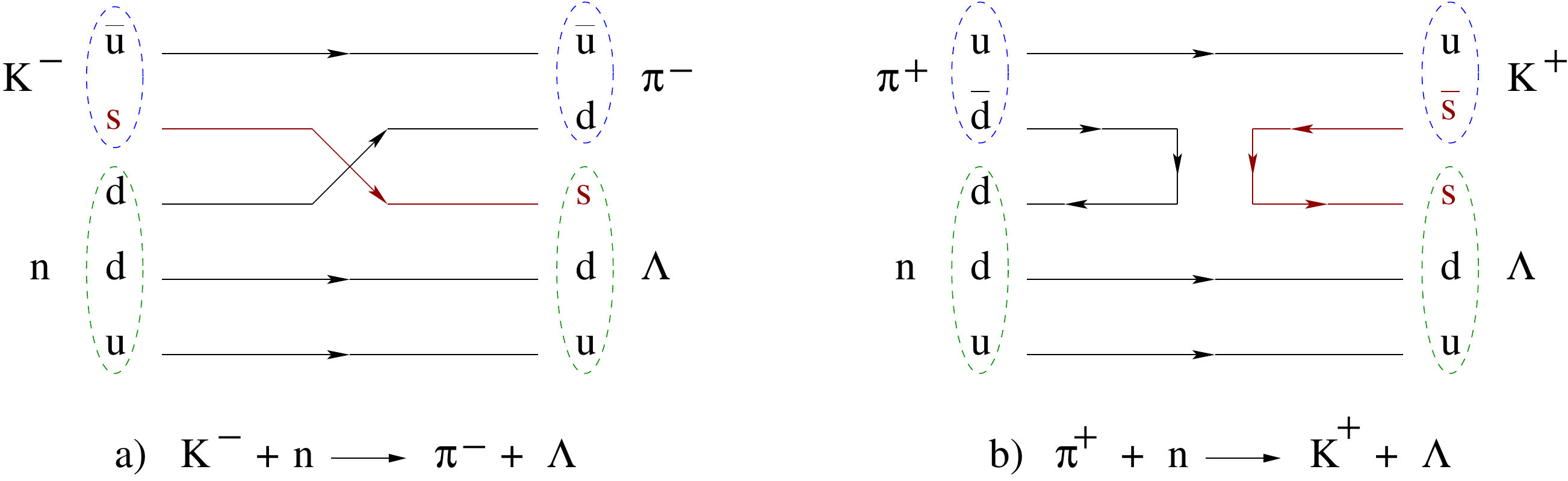} 
\end{center}
\caption{Quark flow diagrams of the elementary processes $K^-+n\rightarrow \pi^-+\Lambda$ (diagram a) and $\pi^++n\rightarrow K^++\Lambda$ (diagram b)  leading the formation of single $\Lambda$-hypernucle in {\it strangeness exchange} and {\it associated production} reactions.} 
\label{fig:elp}
\end{figure}

Hypernuclei can be produced by several mechanisms. The use of separated $K^-$ beams has allowed for instance to produce single $\Lambda$-hypernuclei through $(K^-,\pi^-)$ {\it strangeness exchange reactions:}
\begin{equation}
K^-\,\, + \,\, ^AZ \,\, \rightarrow\,\,  ^A_{\Lambda}Z\,\,+\,\,\pi^- \ ,
\end{equation}
where a $K^-$ hits a neutron of the nuclear target changing it into a $\Lambda$ and emitting a $\pi^-$. The quark flow diagram of the corresponding elementary process $K^-+n\rightarrow \pi^-+\Lambda$, where one of the $d$ quarks of the neutron is exchanged by the $s$ quark of the $K^-$, is shown in Fig.\ \ref{fig:elp} (diagram a). By measuring the momenta of both the incoming $K^-$ and the outgoing $\pi^-$ using two magnetic spectrometers with good energy resolution it is possible to determine accurately the mass of the formed hypernucleus 
\begin{equation}
M({^A_{\Lambda}Z})=\sqrt{\left(E_{\pi^-}-E_{K^-}-M({^AZ})\right)^2+\left(\vec p_{\pi^-}-\vec p_{K^-}\right)^2} \ ,
\end{equation}
from which its binding energy can be easily obtained
\begin{equation}
B({^A_{\Lambda}Z})=B({^AZ})+M({^AZ})+M_{\Lambda}-M_{N}-M({^A_{\Lambda}Z}) \ .
\end{equation}
In some experiments, a rather low-momentum $K^-$ beam is injected on thick nuclear targets. In this case, the $K^-$ is stopped before it decays, losses its energy in the target, and is eventually trapped in an atomic orbit. The stopped $K^-$ is then absorved by the atomic nucleus through a strangeness exchange process that leads to the formation of a hypernucleus and the emission of a $\pi^-$, 
\begin{equation}
K^-_{stopped}\,\, + \,\, ^AZ \,\, \rightarrow\,\,  ^A_{\Lambda}Z\,\,+\,\,\pi^- \ .
\end{equation}
This reaction occurs essentially at rest and, therefore, in this case it is necessary to measure only the momentum of the emitted pion in order to determine the mass of the hypernucleus 
\begin{equation}
M({^A_{\Lambda}Z})=\sqrt{\left(E_{\pi^-}-E_{K^-}-M({^AZ})\right)^2+\vec p_{\pi^-}^2} \ ,
\end{equation}
and its corresponding binding energy. Therefore, only one magnetic spectrometer is necessary in this case. These reactions, initially carried out at CERN, have been studied mainly at BNL in the USA, and at KEK and J-PARC in Japan.

The use of $\pi^+$ beams has permitted to produce hypernuclei by means of $(\pi^+,K^+)$ {\it associated production reactions:}
\begin{equation}
\pi^+\,\, + \,\, ^AZ \,\, \rightarrow\,\,  ^A_{\Lambda}Z\,\,+\,\,K^+ \ .
\end{equation}
In this case, when a neutron of the target is hit by a $\pi^+$, an $s\bar{}s$ pair is created from the vacuum, and a $K^+$ and a $\Lambda$ are produced in the final state. Diagram b of Fig.\ \ref{fig:elp}  shows
the quark flow diagram of the underlying elementary process $\pi^++n\rightarrow K^++\Lambda$. The production cross section of this mechanism is reduced compared to the one of the strangeness exchange reaction. This drawback, however, is compensated by the fact that the intensities of the $\pi^+$
beams are larger than those of the $K^-$ ones. The mass of the hypernucleus and, therefore, its binding energy is obtained by measuring the momenta of the incident $\pi^+$ and the outgoing $K^+$ with the help of two spectrometers as in the case of the $(K^-,\pi^-)$ reaction. These experiments have been also performed at BNL and KEK, and latter at GSI (Germany).

\begin{figure}[t!]
\begin{center}
\includegraphics[width=12cm]{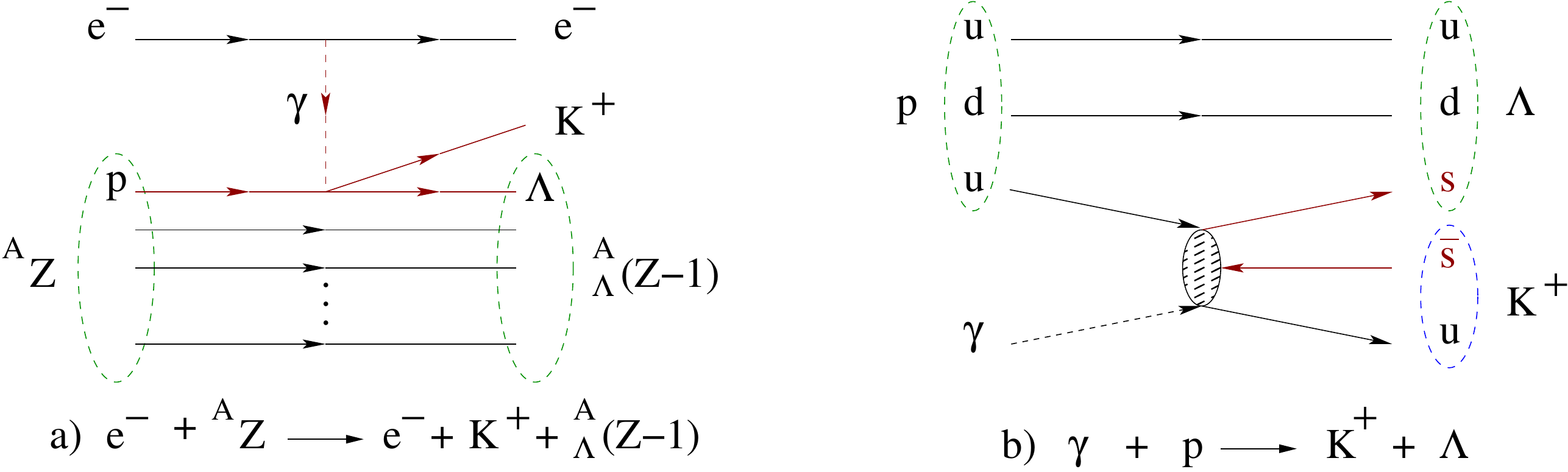} 
\end{center}
\caption{Production of a single $\Lambda$-hypernuclei through the $(e,e'K^+)$ reaction (diagram a) and quark flow diagram of the corresponding elementary process $\gamma+p\rightarrow K^++\Lambda$ (diagram b).} 
\label{fig:elp2}
\end{figure}

The {\it electroproduction} of hypernuclei by means of the $(e,e'K^+)$ reaction, 
\begin{equation}
e^-\,\, + \,\, ^AZ \,\, \rightarrow\,\, e^-\,\,+\,\, K^+\,\,+\,\,  ^A_{\Lambda}(Z-1) \ ,
\end{equation}
provides a high-precision tool for the study of of $\Lambda$-hypernuclear spectroscopy due to the excellent spatial and energy resolution of the electron beams \cite{hugenford94}. This reaction can described in good approximation as the exchange of  a virtual photon between the incoming electron and a proton of the nuclear target (see Fig.\ \ref{fig:elp2}). The electron is scattered and a $\Lambda$ plus a $K^+$ are produced in the final state . The cross section for this reaction is about 2 orders of magnitude smaller than that of the $(\pi^+,K^+)$ one. However, this can be compensated by the largest intensities of the electron beams. 
Experiments must be done within a small angle around the direction of the virtual photon because the cross section falls rapidly with increasing transfer momentum, and the virtual photon flux is maximized for an electron scattering angle near zero degrees. The geometry of the experiment requieres the use of a couple of spectrometers to detect the kaon and the scattered electrons (which define the virtual photon). These spectrometers must be placed at extremely forward angles, making necessary the use of a magnet to deflect the electrons away from zero degrees into their respective spectrometer. Additionally, since many, protons, positrons and pions are transmitted through the kaon spectrometer, an excellent particle identification is requiered, not only in the hardware trigger, but also in the data analysis. By measuring the type of out-going particles and their energies $(E_{e'}, E_{K^+})$, and knowing the energy of the in-coming electron $(E_e)$, it is possible to calculate the energy which is left inside the nucleus in each event:  
\begin{equation}
E_x=E_e-E_{e'}-E_{k^+} \ ,
\end{equation}
from which the binding energy of the produced hypernuclei can be deduced. At the present moment only two laboratories in the world, the JLAB in the USA and MAMI-C in Germany, have the instrumental capabilities to perform experiments on hypernuclear spectroscopy by using electron beams. 

\begin{figure}[t!]
\begin{center}
\includegraphics[width=7cm]{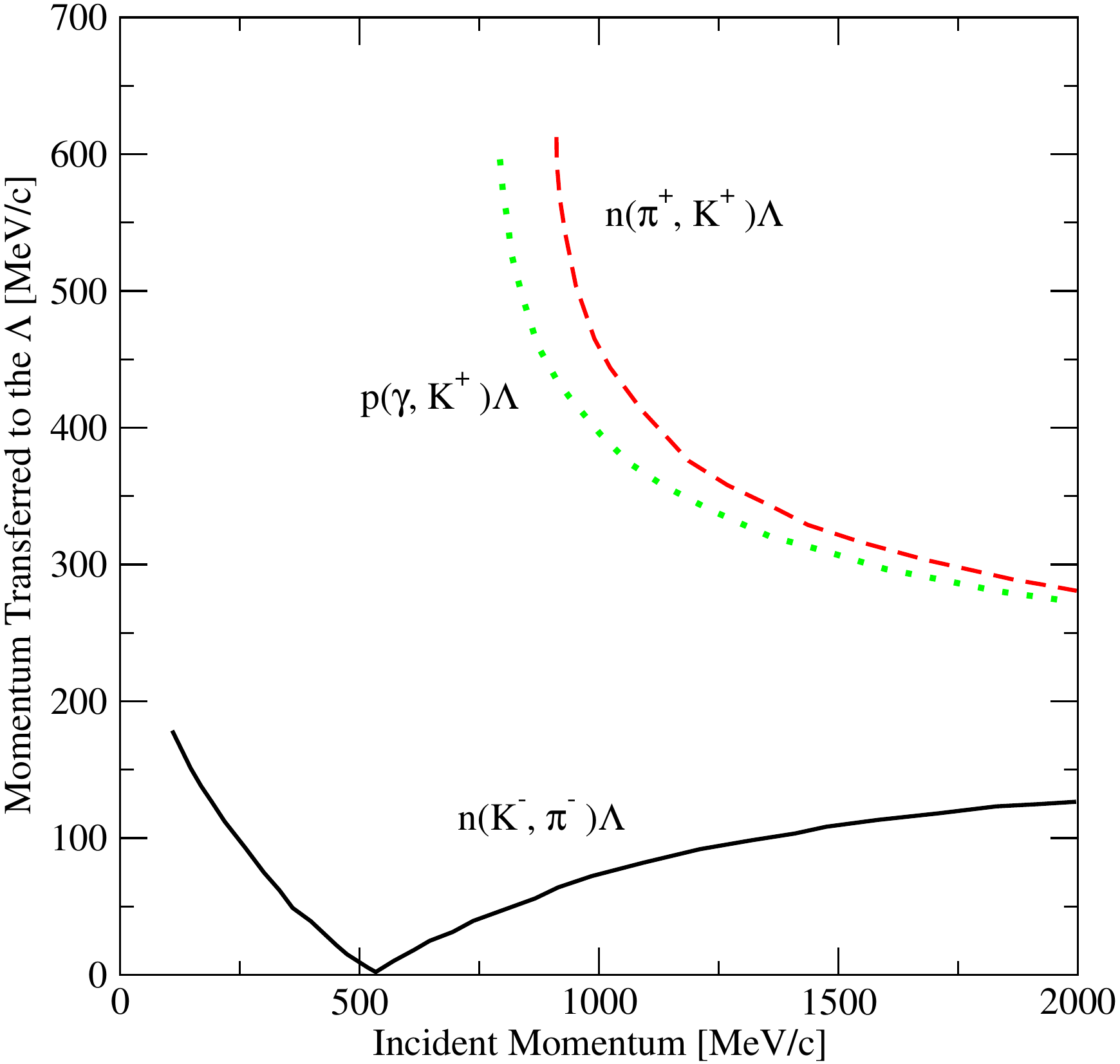} 
\end{center}
\caption{Momentum transferred to the $\Lambda$ as a function of the incident particle momentum for the $n(K^-,\pi^-)\Lambda$, $n(\pi^+,K^+)\Lambda$  and $p(\gamma, K^+)\Lambda$ elementary process at $0^0$ underlying the production of single $\Lambda$-hypernuclei. Figure adapted from Ref.\ \cite{hugenford07}.}
\label{fig:kinematic}
\end{figure}

Before continuing, we should mention here that the HypHI collaboration at FAIR/GSI has recently proposed a completely new and alternative way to produce hypernuclei by using stable and unstable heavy ion beams \cite{hypHI}. A first experiment has been already performed using a $^6$Li beam on a $^{12}$C target at 2 A GeV, in which the $\Lambda$ and the $^3_{\Lambda}$H and $^4_{\Lambda}$H hypernuclei have been observed \cite{rappold13}.

Fig.\ \ref{fig:kinematic} shows the the kinematics of the elementary processes $n(K^-,\pi^-)\Lambda$, $n(\pi^+,K^+)\Lambda$  and $p(\gamma, K^+)\Lambda$ underlying the three production mechanisms of
single $\Lambda$-hypernuclei discussed above. Note that the momentum transferred to the $\Lambda$ is much lower for the first of these reactions than for the other two. This is basically due to the fact that since
the $K^-$ interacts strongly with the nucleus through various resonant states, like {\it e.g.,} the famous $\Lambda$(1405), the in-coming kaons in the $n(K^-,\pi^-)\Lambda$ reaction slows down rapidly in the nucleus, and they interact (with very little momentum transfer) mostly with an outer shell neutron  that is replaced by a $\Lambda$ in the same shell. Consequently, in this case, the $\Lambda$ will have a large probability of interacting with, or being bound to, the nucleus. On the contrary, the mean free path of $\pi^+$ and $K^+$ in the nuclear medium is longer than that of the $K^-$ and, therefore, they can interact with more internal nucleons transferring a larger momentum to the $\Lambda$. Thus, in the case of the $n(\pi^+,K^+)\Lambda$ or $p(\gamma, K^+)\Lambda$ reactions, being the recoil momentum of the hyperon high, the cross sections to bound states are reduced, and the produced $\Lambda$ has a high probability of escaping the nucleus. 

$\Sigma$-hypernuclei can also be produced by the mechanisms just described. However, as mentioned before, there is not yet an unambiguous experimental confirmation of their existence. 

Double-$\Lambda$ hypernuclei are nowadays the best systems to investigate the properties of the strangeness $S=-2$ baryon-baryon interaction. Contrary to single $\Lambda$-hypernuclei, double-$\Lambda$ hypernuclei cannot be produced in a single reaction. To produce them, first it is needed to create a $\Xi^-$ which can be done through reactions like
\begin{equation}
K^-\,\,+\,\,p\,\,\rightarrow\,\,\Xi^-\,\,+\,\,K^+ \ ,
\label{eq:xi1}
\end{equation}
or
\begin{equation}
p\,\,+\,\,\bar p\,\,\rightarrow\,\,\Xi^-\,\,+\,\,\bar \Xi^+ \ .
\label{eq:xi2}
\end{equation}
Then if the $\Xi^-$ is captured in an atomic orbit it can interact in a second step with the nuclear core producing two $\Lambda$ hyperons via proceses such as {\it e.g.,}
\begin{equation}
\Xi^-\,\,+\,\,p\,\,\rightarrow\,\,\Lambda\,\,+\,\,\Lambda\,\,+\,\,28.5\,\,\mbox{MeV}  \ ,
\end{equation}
where the approximatelly $28-29$ MeV of energy reliased in this reaction are equally shared between the two $\Lambda$'s leading, in most cases, to the escape of one or both of them from the nucleus. The bond energy $\Delta B_{\Lambda\Lambda}$ of two $\Lambda$'s in double-$\Lambda$ hypernuclei can  be determined experimentally from the measurement of the biding energies of single- and double-$\Lambda$ hypernuclei simply as
\begin{equation}
\Delta B_{\Lambda\Lambda}=B_{\Lambda\Lambda}(^A_{\Lambda\Lambda}Z)-2B_{\Lambda}(^{A-1}_{\Lambda}Z) \ .
\end{equation}

Earlier emulsion experiments  reported the formation of a few double-$\Lambda$ hypernuclei: $^6_{\Lambda\Lambda}$He, $^{10}_{\Lambda\Lambda}$Be and $^{13}_{\Lambda\Lambda}$B \cite{dl1,dl2,dl3,dl4,dl5}. The subsequent analysis of these experiments indicated a quite large $\Lambda\Lambda$ bond energy of around $4-5$ MeV, contrary to SU(3) expectations \cite{nij1,nij2,nijmegenb,nij3,nij4}.
However, the identification of some of these double-$\Lambda$ hypernuclei was ambiguous and, therefore,
careful attention should be paid to these old data specially when using it to put any kind of constraint on the $\Lambda\Lambda$ interaction. In 2001 a new $^6_{\Lambda\Lambda}$He candidate was unambiguosly observed at KEK \cite{nagara}. The value  of the $\Lambda\Lambda$ bond energy
deduced from this event was $\Delta B_{\Lambda\Lambda}=1.01\pm0.2^{+0.18}_{-0.11}$ MeV which has been recently revised to $\Delta B_{\Lambda\Lambda}=0.67\pm 0.17$ MeV due to a change in the value of the $\Xi^-$ mass \cite{ahn}. Further experiments are planned in the future at BNL, KEK and J-PARC with $K^-$ beams, and at FAIR/GSI with protons and antiprotons.

Finally, we note that $\Xi^-$-hypernuclei can be produced through the reactions (\ref{eq:xi1}) and (\ref{eq:xi2}). As mentioned in the introduction very few $\Xi$-hypernuclei have been presently identified. The analysis of the experimental data from the production reactions $^{12}$C$(K^-,K^+)^{12}_{\Xi^-}$Be reported in Ref.\ \cite{kau} seems to indicate an attractive $\Xi$-nucleus interaction of the order of about $-14$ MeV. Here we should mention also the very recent
observation of a deeply bound state of the $\Xi^--^{14}$N system with a  binding energy of $4.38\pm0.25$ MeV by Nakazawa {\it et al.,} \cite{naka}. This event provides the first clear evidence of a deeply bound state of this system by an attracive $\Xi$N interaction. Future $\Xi$-hypernuclei production experiments are being planned at J-PARC.


\begin{figure}[t!]
\begin{center}
\includegraphics[width=8cm]{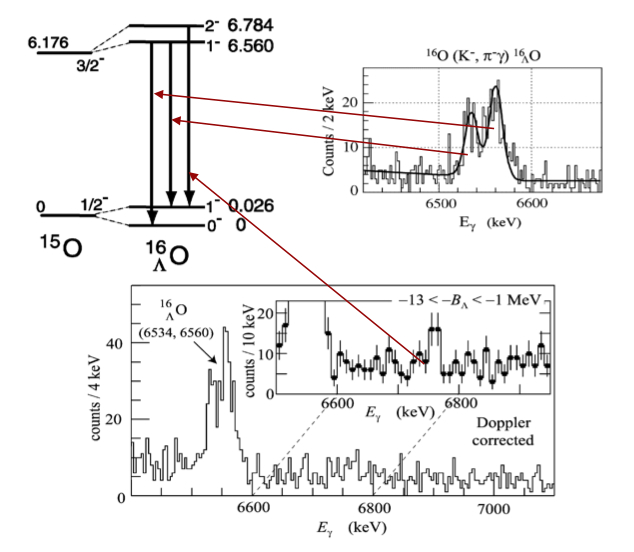} 
\end{center}
\caption{$\gamma$-ray transitions and level scheme of the $^{16}_\Lambda$O measured at BNL. Figure adapted from Ref.\ \cite{ukai08}.}
\label{fig:gamma_ray}
\end{figure}

\subsection{Hypernuclear $\gamma$-ray spectroscopy}

Hypernuclei can be produced in excited states if a nucleon in a $p$ or a higher shell is replaced by a hyperon.  
The energy of these excited states can be released either by emitting nucleons, or, sometimes, when the hyperon moves to lower energy states, by the emission of $\gamma$-rays. The detection of $\gamma$-ray transitions in 
$\Lambda$-hypernuclei has allowed the analysis of hypernuclear excited states with very good energy resolution. However, there have been some technical difficulties in the application of $\gamma$-ray spectroscopy to hypernuclei mainly related with the detection efficiency of $\gamma$-ray measurements
and with the necessity of covering a large solid angle with $\gamma$-ray detectors. 
The construction of  the large-acceptance germanium detector array {\it Hyperball} \cite{hashimoto06}, dedicated to hypernuclear $\gamma$-ray spectroscopy, has allowed to solve somehow these issues.
There exist still, however, several weak points in hypernuclear $\gamma$-ray spectroscopy. A number of single-particle $\Lambda$ orbits are bound in heavy $\Lambda$  hypernuclei with a potential depth of around 30 MeV but the energy levels of many single-particle orbits are above the neutron and proton emission thresholds. Therefore, the observation of $\gamma$-rays is limited to the low excitation region, maybe up to the $\Lambda$ p-shell. The fact that $\gamma$-ray transition only measures the energy difference between two states is clearly another weak point, since single energy information is not enough to fully identify the two levels. The measurement of two $\gamma$-rays in coincidende might help to resolve it. 

Figure \ref{fig:gamma_ray} shows, as an example, the $\gamma$-ray transitions and the level scheme of $^{16}_\Lambda$O identified and determined by $\gamma$-ray spectroscopy using the germanium detector array {\it Hyperball} at BNL \cite{ukai08}. The $\gamma$-ray spectrum of $^{16}_\Lambda$O was measured by using the $(K^-,\pi^-)$ reaction. The observed twin peaks demonstrate the hypernuclear fine structure for the $(1^-\rightarrow 1^-)$ and $(1^-\rightarrow 0^-)$ transitions in $^{16}_\Lambda$O. The small spacing between the twin peaks is due to the spin dependence of the $\Lambda$ N interaction. 


\subsection{Weak decay of single $\Lambda$-hypernuclei}

The main decay mode of a $\Lambda$ particle in free space is the so-called mesonic weak decay mode
\begin{equation}
\Lambda \rightarrow N+\pi \ , \,\,\,\,\,\,\, p_N\sim 100 \,\, \mbox{MeV/c}
\end{equation}
where a $\Lambda$ particle decays $\sim 60\%$ of the times into a proton and a $\pi^-$, and $\sim 40\%$ of them into a neutron and a $\pi^0$. This mode, however, is strongly suppressed by the Pauli principle when the hyperon is bound in the nucleus, because the momentum of the out-going nucleon ($\sim 100$ MeV/c) is smaller than the typical Fermi momentum in the nucleus ($\sim 270$ MeV/c). The so-called non-mesonic mode, according to which the $\Lambda$ interacts with one (or more) of the surrounding nucleons
\begin{equation}
\Lambda+N \rightarrow N+N \ , \,\,\,\,\,\,\, p_N\sim 420 \,\, \mbox{MeV/c} 
\label{eq:g1}
\end{equation}
\begin{equation}
\Lambda+N+N \rightarrow N+N+N \ , \,\,\,\,\,\,\, p_N\sim 340 \,\, \mbox{MeV/c} 
\label{eq:g2}
\end{equation}
becomes, therefore, the dominant decay mode in hypernuclei, specially in medium and heavy hypernuclei. The weak decay of hypernuclei has been mainly studied within the frameworks of meson-exhange models \cite{wdme1,wdme2} and effective field theory \cite{wdeft1,wdeft2}. Two comprehensive reviews on the theoretical aspects of hypernuclear weak decay can be found in Refs.\ \cite{gianni,assumlec} for the interested
reader. 

\begin{figure}[t!]
\begin{center}
\includegraphics[width=7cm]{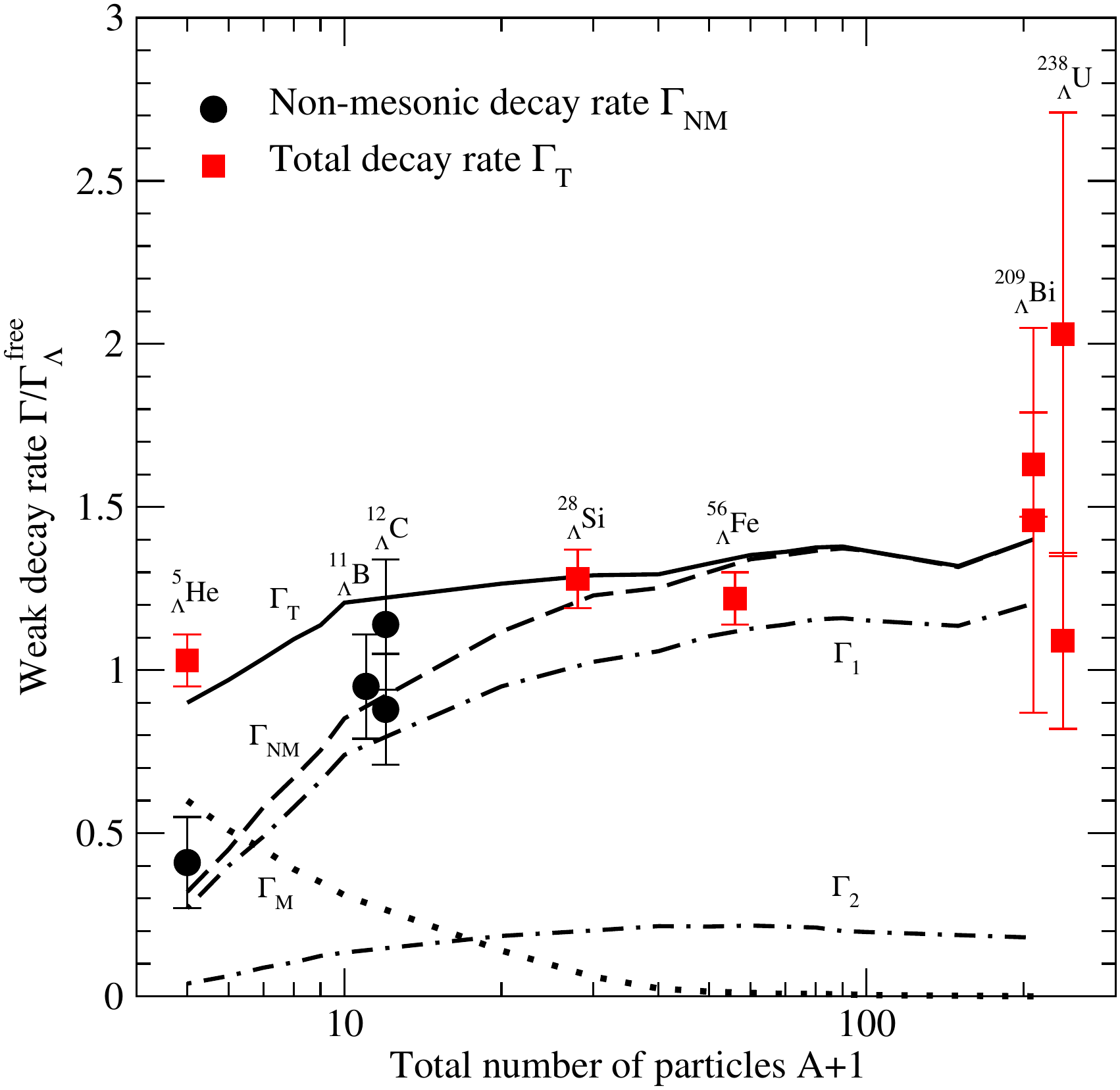} 
\end{center}
\caption{Weak decay rate $\Gamma$ as a function of the total number of particles in units of the weak decay rate  of the $\Lambda$ in free space $\Gamma_\Lambda^{\mbox{free}}$. 
Figure adapted from the original one in Ref.\ \cite{alberico00}.}
\label{fig:gamma}
\end{figure}

The weak decay rate $\Gamma$ (expressed in units of the decay rate of the $\Lambda$ in free space) is shown as function of the total number of particles $A+1$ in Fig.\ \ref{fig:gamma}. The figure has been adapted from the original one in Ref.\ \cite{alberico00}. Theoretical predictions of the mesonic $\Gamma_M$, non-mesonic $\Gamma_{NM}$ and total $\Gamma_T$ decay rates are presented by the dot, dashed and solid lines, respectively. The contributions of one-nucleon and two-nucleon induced decay mode to the non-mesonic decay rate (see Eqs.\ (\ref{eq:g1}) and (\ref{eq:g2})) are also plotted (curves labelled $\Gamma_1$ and $\Gamma_2$ in the figure). Experimental values of the total and non-mesonic decay rates are given by the squares and circle marks respectively. As it can be seen in the figure, the analysis of hypernuclear lifetimes as a function of the mass number A shows that the mesonic decay mode gets blocked as A increases, while the non-mesonic decay increases up to a saturation value of the order of the free decay, reflecting the short-range nature of the weak $\Delta S=1$ baryon-baryon interaction. The interested reader is referred to Refs.\ \cite{wdme1,wdme2,wdeft1,wdeft2,alberico00,botta,feliciello,gianni,assumlec} and references therein for a detailed discussion on this topic.


\section{The hyperon-nucleon interaction}
\label{sec:models2}

Quantum chromodynamics (QCD) is commonly recognized as the fundamental theory of the strong interaction, and therefore, in principle, the baryon-baryon interaction could be completely determined by the underlying quark-gluon dynamics in QCD. Nevertheless, due to the mathematical problems raised by the non-perturbative character of QCD at low and intermediate energies (at this range of energies the coupling constants become too large for perturbative approaches), one is still far from a quantitative undertanding of the baryon-baryon interaction from the QCD point of view. This problem is, however, usually circumvented by introducing a simplified model in which only hadronic degrees of freedom are assumed to be relevant.
Quarks are confined inside the hadrons by the strong interaction and the baryon-baryon force arises from 
meson-exchange \cite{nagels73,machleidt87,na78,nij1,nij2,nijmegenb,nij3,nij4, juelich89,juelich04}. Such an effective description is presently the most quantitative representation of the fundamental theory in the energy regime of nuclear physics, although a big effort is being invested recently in understanding the baryon-baryon interaction from an effective field theory perspective \cite{kolck99}. Quark degrees of freedom are expected to be important only at very short distances and high energies. Short-range parts of the interaction are treated, in all meson exchange model and effective field theory approaches, by including form factors which take into account, in an effective way, the extended structure of hadrons. 

In this section we will briefly review the meson exchange and chiral effective field theory approaches of the hyperon-nucleon and hyperon-hyperon interactions, and we will present recent developments from the the so-called V$_{low\,\, k}$ approach and lattice QCD.


\subsection{Meson exchange models}

\begin{figure}[t!]
\begin{center}
\includegraphics[width=10cm]{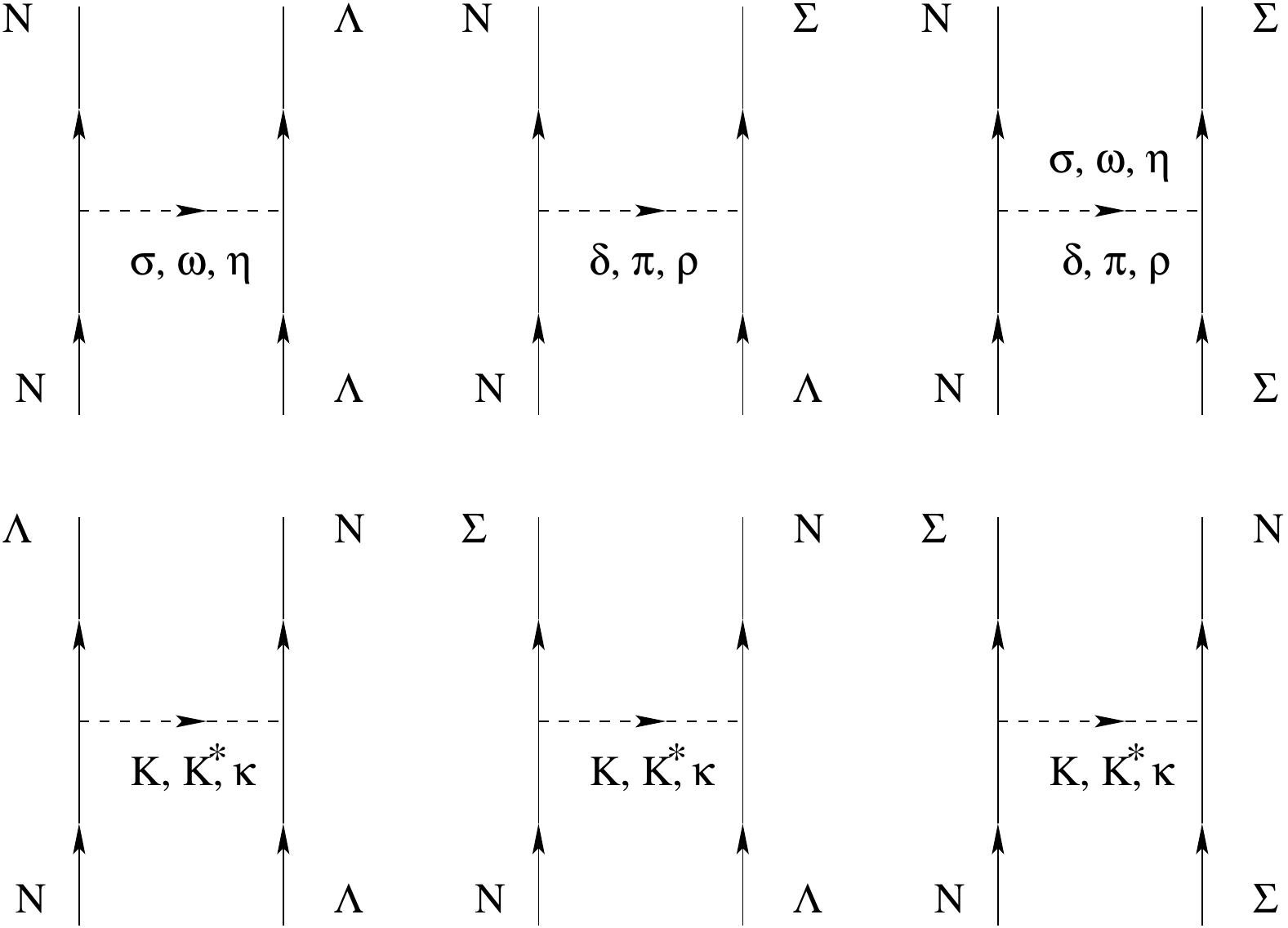} 
\end{center}
\caption{One meson exchange diagrams contributing to the $\Lambda N\rightarrow \Lambda N$,
$\Sigma N \rightarrow \Sigma N$ and $\Lambda N \rightarrow \Lambda N$ interaction channels. }
\label{fig:obe}
\end{figure}

The three relevant meson field types that mediate the interaction among the different baryons are: the scalar (s) fields: $\sigma, \delta$; the pseudoscalar (ps) fields: $\pi, K,\eta,\eta'$; and the vector (v) fields: $\rho,K^*,\omega,\phi$. Guided by symmetry principles, simplicity and physical intuition the most commonly employed interaction Lagrangians that couple these meson fields to the baryon ones are
\begin{eqnarray}
{\mathcal{L}}_s&=&g_s\bar \Psi\Psi \Phi^{(s)} \\
{\mathcal{L}}_{ps}&=&g_{ps}\bar \Psi i\gamma^5\Psi \Phi^{(ps)} \\
{\mathcal{L}}_v&=&g_v\bar \Psi \gamma^\mu\Psi \Phi^{(v)_\mu}+g_t\bar \Psi \sigma^{\mu\nu}\Psi
\left( \partial_\mu\Phi^{(v)}_nu-\partial_\nu\Phi^{(v)}_\mu\right) 
\end{eqnarray}
for scalar, pseudoscalar and vector coupling, respectively. Alternatively, for the pseudoscalar field there is also the so-called pseudovector (pv) or gradient coupling, which is suggested as an effective coupling by chiral symmetry \cite{weinberg67,brown79}
\begin{equation}
{\mathcal{L}}_{pv}=g_{pv}\bar \Psi \gamma^5\gamma^\mu\Psi\partial_\mu\Phi^{(ps)} \ .
\end{equation}
In the above expressions $\Psi$ denotes the baryon fields for spin $1/2$ baryons, $\Phi^{(s)}, \Phi^{(ps}$ and $\Phi^{(v)}$ are the corresponding scalar, pseudoscalar and vector fields, and the g's are the corresponding coupling constants that must be constrained by {\it e.g.} scattering data. Note that the above Lagrangians are for isoscalar mesons, however, for isovector mesons, the fields $\Phi$ trivially modify to $\vec \tau\cdot \vec \Phi$ with $\vec \tau$ being the usual isospin Pauli matrices.

Employing the above Lagrangians, it is possible to construct a one-meson-exchange (OME) potential model. A typical contribution to the baryon-baryon scattering amplitude arising from the exchange of a certain meson $\Phi$ is given by
\begin{equation}
\langle p_1'p_2'|V_\Phi|p_1p_2\rangle =
\frac{\bar u (p_1')g_{\Phi 1}\Gamma_\Phi^{(1)}u(p_1)P_\Phi \bar u (p_2')g_{\Phi 2}\Gamma_\Phi^{(2)}u(p_2)}{(p_1-p_1')^2-m_\Phi^2}
\end{equation}
where $m_\Phi$ is the mass of the exchanged meson, $P_\Phi/((p_1-p_1')^2-m_\Phi^2)$ represents the 
meson propagator, $u$ and $\bar u$ are the usual Dirac spinor and its adjoint ($\bar uu=1, \bar u=u^\dagger\gamma^0$), $g_{\Phi 1}$ and $g_{\Phi 2}$ are the coupling constants at the vertices, and the $\Gamma$'s denote the corresponding Dirac structures of the vertices
\begin{eqnarray}
\Gamma_s^{(i)}=1 \ , \,\, 
\Gamma_{ps}^{(i)}=i\gamma^5 \ , \,\,
\Gamma_{v}^{(i)}=\gamma^\mu \ , \,\,
\Gamma_{t}^{(i)}=\sigma^{\mu\nu} \ , \,\,
\Gamma_{pv}^{(i)}=\gamma^5\gamma^\mu\partial_\mu \ .
\end{eqnarray}

In the case of scalar and pseudoscalar meson-exchanges, the numerator $P_\Phi$ of the propagator is just 1. For vector meson-exchange, however, is the rank 2 tensor
\begin{equation}
P_\Phi\equiv P_{\mu\nu}= -g_{\mu\nu}+\frac{q_\mu q_\nu}{m_\Phi^2} \ ,
\label{eq:numerator}
\end{equation}
where $g_{\mu\nu}=$diag(1,-1,-1,-1) is the usual Minkowski metric tensor and $q_\mu=(p_1-p_1')_{\mu}$ is the four momentum transfer.

In general, when all types of baryons are included, the scattering amplitue will be simply the sum of all the partial contributions
\begin{equation}
\langle p_1'p_2'|V|p_1p_2\rangle = \sum_{\Phi} \langle p_1'p_2'|V_{\Phi}|p_1p_2\rangle \ .
\end{equation}

Expanding the free Dirac spinor in terms of 1/M (M is the mass of the relevant baryon) to lowest order leads to the familiar non-relativistic expressions for the baryon-baryon potentials, which through Fourier transformation give the configuration space version of the interaction. The general expression for the local approximation of the baryon-baryon interaction in configuration space is
\begin{eqnarray}
V(\vec r) = \sum_{\Phi}\left\{
C_{C_\Phi} + C_{\sigma_\Phi}\vec \sigma_1\cdot \vec \sigma_2
+C_{LS_\Phi}\left(\frac{1}{m_\Phi r}+\frac{1}{(m_\Phi r)^2} \right)\vec L \cdot \vec S \right.\nonumber \\
\left. +C_{T_\Phi}\left(1+\frac{3}{m_\Phi r} + \frac{3}{(m_\Phi r)^2} \right)S_{12}(\hat r)
\right\}
\frac{e^{-m_\Phi r}}{r} , 
\label{eq:rspace}
\end{eqnarray}
where $C_{C_\Phi}, C_{\sigma_\Phi}, C_{LS_\Phi}$ and $C_{T_\Phi}$ are numerical factors containing the coupling constants g's and the baryon masses, $\vec L$ is the total orbital angular momentum, $\vec S$ is the total spin, and $S_{12}(\hat r)$ is the usual tensor operator in configuration space,
\begin{equation}
S_{12}(\hat r)=3(\vec \sigma_1\cdot \hat r)(\vec \sigma_2 \cdot \hat r) - (\vec \sigma_1\cdot \sigma_2) \ , \,\,\, \hat r=\frac{\vec r}{|\vec r|} \ .
\label{eq:tensor}
\end{equation}

Finally, one has to remember that in the meson exchange theory all meson-baryon vertices must be necessarily modified by the introduction of the so-called form factors. Each vertex is multiplied by a form factor of the type
\begin{equation}
F_\alpha(|\vec k|^2)=\left(\frac{\Lambda_\alpha^2-m_\alpha^2}{\Lambda_\alpha^2+|\vec k|^2} \right)^{n_\alpha}
\label{eq:ff}
\end{equation}
or by
\begin{equation}
F_\alpha(|\vec k|^2)=\mbox{exp}\left(-\frac{|\vec k|^2}{2\Lambda_\alpha^2}\right) \ .
\label{eq:ff2}
\end{equation}
In Eq.\ (\ref{eq:ff}) the quantity $n_\alpha$ is usualy taken equal to 1 (monopole form factor) or 2 (dipole form factor). The vector $\vec k$ denotes the 3-momentum transfer, whereas $\Lambda_\alpha$ is the so-called cut-off mass, typically of the order 1.2 - 2 GeV. Originally the form factors were introduced for purely mathematical reasons, namely, to avoid divergences in the scattering equation. Nevertheless, our present knowledge of the (quark) substructure of baryons and mesons provides a physical reason for their presence. Obviously, it does not make sense to take the meson exchange picture seriously in a region in which modifications due to the extended structure of hadrons come into play.

Until now all that we have said is general and nothing has been commented yet about the specific hyperon-nucleon. Presently there are in the market two different meson exchange models for the hyperon-nucleon: the J\"{u}lich models \cite{juelich89,juelich04} and the Nijmegen \cite{nij1,nij2,nijmegenb,nij3,nij4} ones. The main features of these two models are briefly presented in the following and the interested reader is referred to the original works for detailed information.

The J\"{u}lich hyperon-nucleon interaction \cite{juelich89,juelich04} is constructed in complete analogy to the Bonn nucleon-nucleon force \cite{machleidt87}. It is defined in momentum space and contains the full energy-dependence and non-locality structure. Besides single-meson exchange processes, it includes higher-order processes involving 
$\pi$- and $\rho$-exchange processes (correlated 2$\pi$-exchange are conveniently parametrized in terms of an effective $\sigma$-exchange) and, in adition, $KK$, $KK^*$ and $K^*K^*$ processes with N, $\Delta$, $\Lambda$, $\Sigma$ and $\Sigma^*$(1385) intermediate states. Therefore, the model not only includes the
couplings between the $\Lambda$N and $\Sigma$N channels, but also couplings to the $\Delta\Lambda$, 
$\Delta\Sigma$ and N$\Sigma^*$ ones. The exchange of the pseudoscalar mesons $\eta$ and $\eta'$ is not
consisdered. Parameters (coupling constants and cut-off masses) at NN and N$\Delta$ vertices are taken from the Bonn model. Coupling constants at the vertices involving strange particles are fixed by relating them, under the assumption of SU(6) symmetry, to the NN and N$\Delta$ values. Thus, the only free parameters are the cutt-off masses at the strange vertices which are adjusted to the existing hyperon-nucleon data. The form factors at the vertices are parametrized in the conventional monopole form or dipole form when the vertex involves both a spin-$3/2$ baryon and a vector meson.

The Nijmegen Soft-Core 89 (NSC89) hyperon-nucleon interaction \cite{nij1} is obtained by a straightforward extension of the Nijmegen nucleon-nucleon model \cite{na78}, through the application of SU(3) symmetry. It is defined both in momentum and in configuration space. The model is generated by the exchange of nonets of pseudoscalar and vector mesons, and scalar mesons. Assuming SU(3) symmetry all the coupling constants at the vertices with strange particles are related to the NN ones. Gaussian form factors are taken at the vertices to guarantee a soft behaviour of the potentials in configuration space at small distances.

Finally, the Nijmegen Soft-Core 97 (NSC97a-f) \cite{nijmegenb,nij2} and the recent Extended Soft-Core (ESC) \cite{nij3,nij4} baryon-baryon interactions for the complete octet of baryons are based on SU(3) extensions of the Nijmegen potentials models for the nucleon-nucleon \cite{na78} and the hyperon-nucleon \cite{nij1} interactions. It describes not only the sectors of strangeness S=0 (NN) and S=-1 ($\Lambda$N, $\Sigma$N), but also the ones of strangeness  S=-2 ($\Lambda\Lambda, \Lambda\Sigma, \Sigma\Sigma, \Xi$N), S=-3 ($\Lambda\Xi, \Sigma\Xi$) and S=-4 ($\Xi\Xi$). It is parametrized in terms of one-boson exchanges, and all counpling constants are determined by a fit to the NN and YN scattering data and the use of SU(3) relations. However, the fit to the NN and YN data still allows for some freedom in the parameters and different models exists. These models are characterized by differenty choices of the magnetic vector $F/(F+D)$ ratio, $\alpha_v^m$, which serves to produce different scattering length in the $\Lambda$N and $\Sigma$N channels, but at the same time allows to describe the available NN and YN scattering data as well. Within each model, there are no free parameters left and so each parameter set defines a baryon-baryon potential that models all possible two-baryon interactions. Gaussian form factors are taken as in the NSC89 model.


\subsection{Chiral effective field theory approach}

Althought the meson-exchange picture provides a practical and systematic approach to the description of hadronic reactions in the low- and medium-energy regime, in the last decade chiral effective field theory 
($\chi$EFT) has emerged as a new powerful tool as already mentioned in the introduction. The derivation of the nuclear force from $\chi$EFT has been extensively discussed in the literature since the pioneering work of Weinberg \cite{wei91,wei91b}. The main advantage of this scheme is that there is an underlying power counting that allows to improve calculations
systematically by going to higher orders in a perturbative expansion. In addition it is possible to derive two- and corresponding three-body forces as well as external current operators in a consistent way. During the last years the NN interaction has been described to high precision using $\chi$EFT \cite{entem03,epelbaum05}. In these works, the power counting is applied to the NN potential, which consists of pion exchanges and a series of contact interactions with an increasing number of derivatives to parametrize the shorter part of the NN force. A regularized Lipmann-Schwinger equation is solved to calculate observable quantities. The interested reader is referred to Refs.\ \cite{entem03,epelbaum05} and references therein for a comprehensive review (see also Refs.\ \cite{xeft1,xeft2,xeft3}).

Compared to the NN case, there are very few investigations of the YN interaction using $\chi$EFT. A recent application of the scheme used in Ref.\ \cite{epelbaum05} to the YN and the YY interactions has been performed by the J\"{u}lich-Bonn-Munich group \cite{polinder06,haidenbauer13}. In the next we present a brief description of this $\chi$EFT approach to the YN interaction and refer the interested reader to the original works of the  J\"{u}lich-Bonn-Munich group for details.

\begin{figure}[t!]
\begin{center}
\includegraphics[width=10cm]{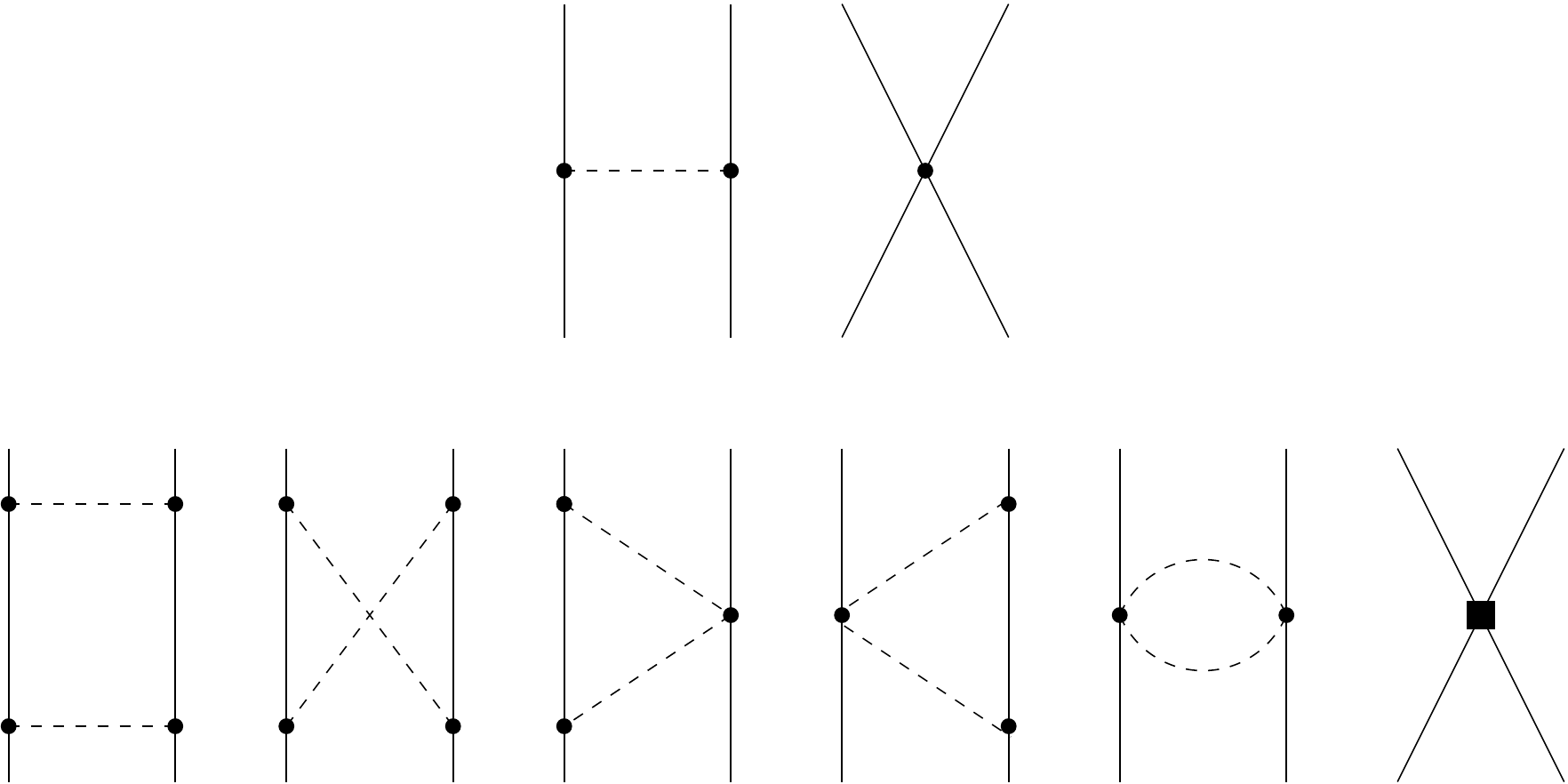} 
\end{center}
\caption{Leading order (upper diagrams) and next-to-leading order (lower diagrams) contributions to the baryon-baryon interaction. Figure adapted from Ref.\ \cite{haidenbauer13}.}
\label{fig:xeft}
\end{figure}

Analogous to the NN potential, at leading order (LO) in the power counting (see the upper diagrams of Fig.\ \ref{fig:xeft}), the YN potential consist of  one pseudoscalar-meson exchanges and of four-baryon contact terms, where each of these two contributions is constrained via SU(3)-flavor symmetry. The contribution from the one pseudoscalar-meson exchange term is constructed from the Lagrangian density
\begin{equation}
{\mathcal{L}}=\langle 
i \bar B \gamma^\mu D_\mu B-M_0\bar B B +\frac{D}{2}\bar B \gamma^\mu\gamma_5\{u_\mu,B\}
+\frac{F}{2}\bar B\gamma^\mu\gamma_5[u_\mu,B]
\rangle \ ,
\label{eq:loopsx}
\end{equation}
where the brackets denote the trace in flavor space, $B$ is the irreducible baryon octet representation of SU(3)$_f$ given by
\begin{eqnarray}
B = \left(
\begin{array}{ccc}
\frac{\Sigma^0}{\sqrt{2}}+\frac{\Lambda}{\sqrt{6}} & \Sigma^+ &  p \\
\Sigma^- &  -\frac{\Sigma^0}{\sqrt{2}}+\frac{\Lambda}{\sqrt{6}} & n \\
 -\Xi^-& \Xi^0 & -\frac{2\Lambda}{\sqrt{6}} 
\end{array}
\right) \ ,
\label{eq:b}
\end{eqnarray}
$D_\mu$ is the covariant derivative, $M_0$ is the octet baryon mass in the chiral limit, $F$ and $D$ are coupling constants satisfying the relation $F+D=g_A\simeq 1.26$ with $g_A$ the axial-vector strength and
$u_\mu=iu^\dag\partial_\mu Uu^\dag$ with
\begin{equation}
U=exp\left(\frac{2iP}{\sqrt{2}F_\pi}\right) \ ,
\label{eq:u}
\end{equation}
being $F_\pi= 92.4$ MeV the weak pion decay constant and
\begin{eqnarray}
P = \left(
\begin{array}{ccc}
\frac{\pi^0}{\sqrt{2}}+\frac{\eta}{\sqrt{6}} & \pi^+ &  K^+ \\
\pi^- &  -\frac{\pi^0}{\sqrt{2}}+\frac{\eta}{\sqrt{6}} & K^0 \\
 -K^-& \bar K^0 & -\frac{2\eta}{\sqrt{6}} 
\end{array}
\right) \ 
\label{eq:p}
\end{eqnarray}
the SU(3)$_f$ irreducible octet representation of the pseudoscalar mesons. The form of the baryon-baryon potentials obtained from this contribution are similar to the ones derived from the meson-exchange approach and in momentum space read
\begin{equation}
V^{BB}_{OBE}=-f_{B_1B2P}f_{B_2B_4P}\frac{(\vec \sigma_1\cdot \vec q)(\vec \sigma_2\cdot\vec q)}{{\vec q}^{\,\,2}+m_{ps}^2 }{\mathcal{I}}_{B_1B_2\rightarrow B_3B_4} \ ,
\label{eq:obe}
\end{equation}
with $f_{B_1B2P}$ and $f_{B_2B_4P}$ the coupling constants of the two vertices, $m_{ps}$ the mass of the exchanged pseudoscalar meson, $\vec q$ the transferred mometum, and ${\mathcal{I}}_{B_1B_2\rightarrow B_3B_4}$ the corresponding isospin factor.

The contribution from the four-baryon contact interactions can be derived from the following minimal
set of Lagrangian densities 
\begin{eqnarray}
{\mathcal{L}}^1=C_i^1\langle \bar B_a\bar B_b (\Gamma_i B)_b(\Gamma_i B)_a\rangle  \nonumber \\
{\mathcal{L}}^2=C_i^2\langle \bar B_a (\Gamma_i B)_a \bar B_b (\Gamma_i B)_b \rangle 
\end{eqnarray}
and
\begin{equation}
{\mathcal{L}}^3=C_i^3\langle \bar B_a (\Gamma_i B)_a\rangle \langle \bar B_b (\Gamma_i B)_b \rangle \ .
\end{equation}
Here, the labels $a$ and $b$ are the Dirac indices of the particles
and $\Gamma_i$ denotes the five elements of the Clifford algebra, $\Gamma_1=1, \Gamma_2=\gamma^\mu, \Gamma_3=\sigma^{\mu\nu}, \Gamma_4=\gamma^\mu\gamma^5, \Gamma_5=\gamma^5$ which are actually diagonal $3\times3$ matrices in the flavor space.  In LO these Lagrangian densities give rise to six independent low-energy coefficients (LECs): $C_S^1,C_T^1,C_S^2,C_T^2,C_S^3$ and $C_T^3$, where S and T refer to the central and spin-spin parts of the potential respectively.  The LO contact potentials for the different baryon-baryon interactions resulting from these Lagrangians have the general form
\begin{equation}
V^{BB}_{L0}=C_S^{BB} + C_T^{BB}\vec \sigma_1\cdot \vec \sigma_2 \ ,
\label{eq:cpot}
\end{equation}
where the coefficients $C_S^{BB}$ and $C_T^{BB}$ are linear combinations of $C_S^1,C_T^1,C_S^2,C_T^2,C_S^3$ and $C_T^3$.

At  next-to-leading order (NLO) the contact  terms read
\begin{eqnarray}
V^{BB}_{NLO}&=&C_1{\vec q}^{\,\,2}+C_2{\vec k}^{\,\,2}
+(C_3{\vec q}^{\,\,2}+C_4{\vec k}^{\,\,2})\vec \sigma_1\cdot\vec\sigma_2
+\frac{i}{2}C_5(\vec\sigma_1+\vec\sigma_2) \cdot (\vec q \times\vec k) \nonumber \\
&+&C_6(\vec q \cdot \vec\sigma_1)(\vec q\cdot\vec\sigma_2)
+C_7(\vec k\cdot\vec\sigma_1)(\vec k\cdot\vec\sigma_2)
+C_8(\vec\sigma_1-\vec\sigma_2) \cdot (\vec q\times\vec k) \ ,
\label{eq:cpot2}
\end{eqnarray}
where $C_i$ $(i=1,\cdots , 8)$ are additional LECs. The momenta $\vec q$ and $\vec k$ are  defined 
in terms of the initial ${\vec p}$ and final ${\vec p\,\,'}$ baryon momenta in the center-of-mass frame as
$\vec q =\vec p\,\,'-\vec p$ and $\vec k =(\vec p+\vec p\,\,')/2$, respectively. The expresions for the two-pseudoscalar meson exchange contributions are rather cumbersome and we refer the interested reader to  the original work of Haidenbauer {\it et al.,} \cite{haidenbauer13} for details.

The baryon-baryon potentials contructed in this way are then inserted in the Lipmann-Schwinger equation which is regularized with a cut-off regulator function of the type
\begin{equation}
F(p,p')=\mbox{exp}\left(-\frac{p^4+p'^4}{\Lambda^4}\right) 
\end{equation}
in order to remove high-energy components of the baryon and pseudoscalar meson fields. The cut-off $\Lambda$ is usually taken in the range $450-700$ MeV.


\subsection{V$_{low\,\, k}$ hyperon-nucleon interaction}

Following the same idea that in the NN case made possible to calculate a "universal" effective low-momentum potential V$_{low\,\, k}$ by using Renormalization Group techniques, recently, Schaefer {\it et al.,} \cite{shaefer06} have generalized this method to the YN sector.
The effective low-momentum potential V$_{low\,\, k}$ is obtained by integrating out the high-momentum components of a realistic YN interaction. This is achieved by introducing a cutoff for the intermediate momenta in the Lipmann-Schwinger equation such that the physical low-energy quantities are cutoff independent. This results in a modified Lipmann-Schwinger equation with a cutoff-dependent effective potential V$_{low\,\, k}$
\begin{equation}
T(k',k;k^2)=V_{low\,\, k}(k',k)+\frac{2}{\pi}P\int_{0}^{\Lambda}dqq^2\frac{V_{low\,\, k}(k',q)T(q,k;k^2)}{k^2-q^2} \ .
\end{equation}
By demanding $dT(k',k;k^2)/d\Lambda=0$, an exact Renormalization Group flow equation for V$_{low\,\, k}$ can be obtained
\begin{equation}
\frac{dV_{low\,\,k}(k',k)}{d\Lambda}=\frac{2}{\pi}\frac{V_{low\,\,k}(k',\Lambda)T(\Lambda,k;\Lambda^2)}{1-k^2/\Lambda^2} \ .
\end{equation}
Integrating this flow equation one can obtain a phase-shift, energy independent, soft ({\it i.e.,} without hard core) and hermitian low-momentum potential V$_{low\,\, k}$. Unfortunatelly, as it has already been said, contrary to the NN case there exist only few YN scattering data and hence the YN interaction is not well constrained. Schaefer 
{\it eit al.,} found (see Figs. 1-6 of Ref.\ \cite{shaefer06}) that the YN phase shifts have approximately the same shape but have different heights, and the diagonal matrix elements, although they collapse for momenta near the cut-off, they differ for lower momenta. In conclusion, however, one can still say that in general the results seem to indicate a similar convergence to an "universal" softer low-momentum YN intertaction as for the NN case. 


\subsection{Baryon-baryon interactions from lattice QCD}

In the recent years a big progress to derive baryon-baryon interactions from lattice QCD has been made by the HALQCD \cite{halqcd,halqcdb,halqcdc} and NPLQCD \cite{nplqcd,nplqcd1,nplqcd2,nplqcd3,nplqcd4} collaborations. Some of their recent results are mentioned here, and we refer the interested reader to the original works of these two collaborations. 

The HALQCD collaboration follows a method to extract the different baryon-baryon potentials from the Nambu--Bethe--Salpeter wave function measured on the lattice. 
Recently this collaboration managed to approach the region of physical masses obtaining results for various nucleon-nucleon, hyperon-nucleon and hyperon-hyperon interaction channels \cite{halqcd1,halqcd2,halqcd3}
at a single value of the lattice volume and of the lattice spacing.

The NPLQCD collaboration combines calculations of correlation functions at several light quark mass values with the low-energy efective field theory. This approach is particularly interesting since it allows to match lattice QCD results with low-energy effective field theories providing the means for first predictions in the physical quark mass limit. 
Recently, this collaboration has calculated the nucleon-nucleon interaction in the $^1S_0$ partial
wave and the $^3S_1-^3D_1$ coupled ones at a pion mass $m_\pi=450$ MeV \cite{nplqcd4}. Although the binding of the $np$ calculated is too large and even the two-neutron system is bound for this pion mass, extrapolations to the physical value of the pion mass indicate that lattice results approach the observed properties of these systems. Very recently the NPLQCD collaboration has also performed lattice QCD calculations of the nuclear matrix elements relevant for  the double-$\beta$ decay $nn\rightarrow ppe^-e^-\bar\nu_e\bar\nu_e$ \cite{nplqcd5}, and the proton-proton fusion cross section $pp\rightarrow de^+\nu$ as well as the Gamow--Teller matrix element contributing to tritium $\beta$-decay \cite{nplqcd6}. In the strangeness sector, this collaboration has been able to determine the binding energies of light hypernuclei including $^3_\Lambda$He, $^4_\Lambda$He and $^4_{\Lambda\Lambda}$He \cite{nplqcd7}; to compute the magnetic moment of the octet baryon \cite{nplqcd8}; and to constraint the interactions of two-baryon octets at the SU(3)-flavor symmetric-point \cite{nplqcd9}. These results have been obtained at pion mass of $\sim 800$ MeV but calculations at $m_\pi = 450$ MeV or lower values, that will allow to extrapolate the results to the physical mass, are in progress.


\section{Hyperons and Neutron Stars}
\label{sec:stars}

Neutron stars are the remnants of the gravitational collapse of massive stars during a Type-II, Ib or Ic
supernova event. Their masses and radii are typically of the order of $1-2 M_\odot$ 
($M_\odot \simeq 2 \times 10^{33}$g being the mass of the Sun) and $10-12$ km, respectively. With 
central densities in the range of $4-8$ times the normal nuclear matter saturation density,
$\epsilon_0 \sim 2.7 \times 10^{14}$ g/cm$^3$ ($\rho_0 \sim 0.16$ fm$^{-3}$), neutron stars are
most likely among the densest objects in the Universe \cite{shapiro,shapiro2,shapiro3}. These objects are an excellent
observatory to test our present understanding of the theory of strong interacting matter at extreme
conditions, and they offer an interesting interplay between nuclear processes and astrophysical observables.

Conditions of matter inside neutron stars are very different from those one can find in Earth, therefore, a
good knowledge of the Equation of State (EoS) of dense matter is required to understand the properties
of these objects. Nowadays, it is still an open question which is the true nature of neutron stars. 
Traditionally the core of neutron stars has been modeled as a uniform fluid of neutron-rich nuclear matter
in equilibrium with respect to the weak interaction ($\beta$-stable matter). Nevertheless, due to the large 
value of the density, new hadronic degrees of freedom are expected to appear in addition to nucleons. 
Hyperons, baryons with a  strangeness content, are an example of these new degrees of freedom. Contrary to terrestial 
conditions, where hyperons are unstable and decay into nucleons through the weak interaction, the equilibrium
conditions in neutron stars can make the inverse process happen. Hyperons may appear in the inner core of neutron stars at densities of 
about $2-3 \rho_0$. Their presence on the neutron star interior leads to a softening of the EoS and cosequently to a reduction of the maximum mass.

Other neutron star 
properties, such as their thermal and structural evolution, can be also very sensitive to the composition, and therefore to the
hyperonic content of neutron star interiors. In particular, the cooling of neutron stars may be affected by the presence of hyperons, 
since they can modify neutrino emissivities and can allow for fast cooling mechanisms.  Furthermore, the emission of gravitational 
waves in hot and rapidly rotating neutron stars due to the so-called r-mode instability can also be affected by the presence 
of hyperons in neutron stars, because  the bulk viscosity of neutron star matter is dominated by the
contribution of hyperons as soon as they appear in the neutron star interior. 

In the following we briefly review the hyperon puzzle and present some of the ideas proposed to solve it. Then we revise the role of hyperons on the properties of newly born neutron stars, neutron star cooling, and  the r-mode instability.


\subsection{The hyperon puzzle}
\label{sec2}

\begin{figure*}[t]
\begin{center}
\resizebox{0.90\textwidth}{!}
{
\includegraphics[width=5cm]{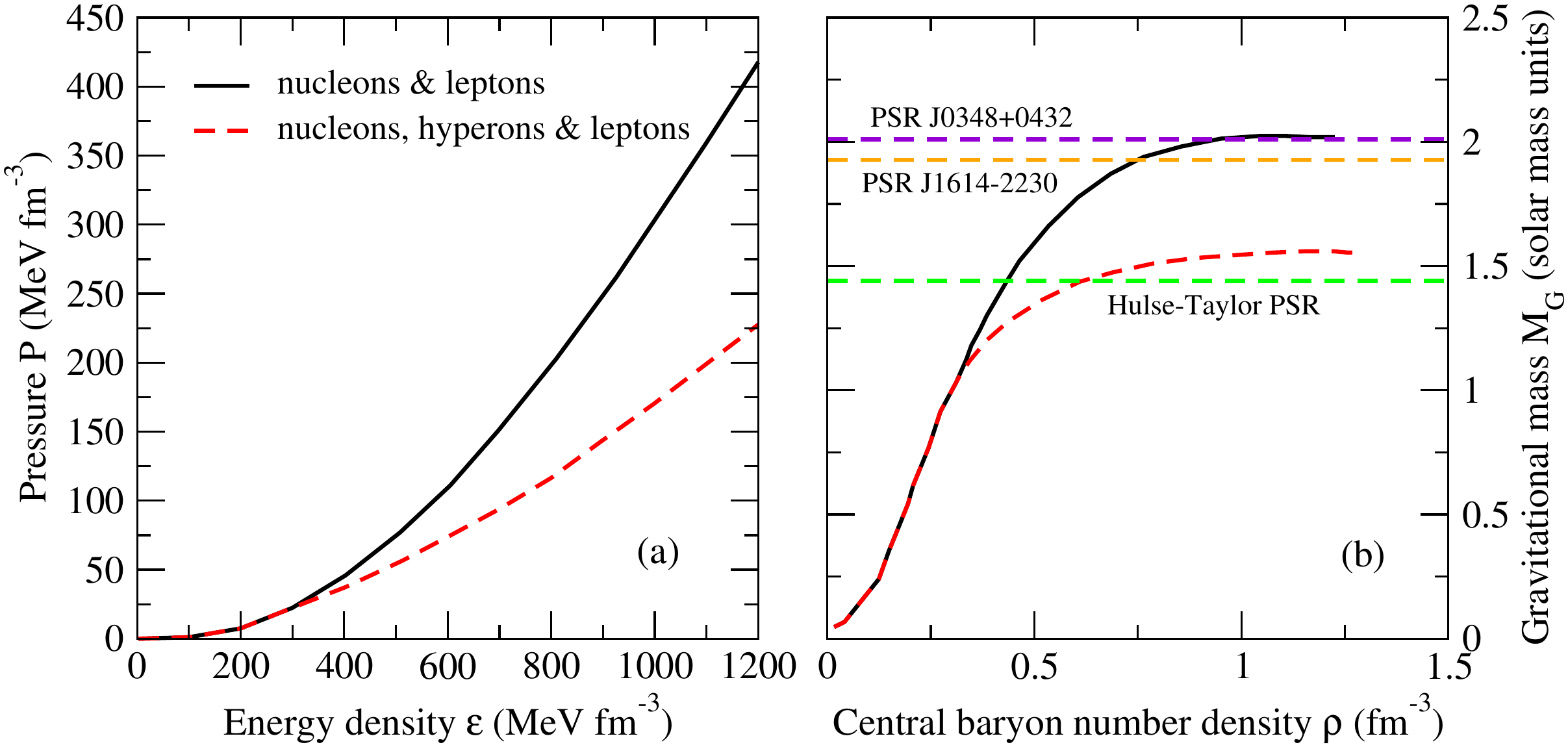}
}
\caption{ Illustration of the effect of the presence of hyperons on the EoS (panel (a)) and mass of a neutron star (panel (b)). A generic model with (black solid line) and without (red dashed line) hyperons has been considered. The horizontal lines shows the observational mass of the Hulse--Taylor \cite{hulsetaylor} pulsar and the recently observed PSR J1614-2230 \cite{demorest} and PSR J0348+0432 \cite{antoniadis}.} 
\label{fig:EOS}       
\end{center}
\end{figure*}

The presence of hyperons in neutron stars was considered for the first time in the pioneering work of Ambartsumyan and Saakyan in 1960 \cite{ambart}. Since then, their effects on the properties of these objects have been studied by many authors using either phenomenological \cite{shf1,shf2,glend,glend2,glend3,glend4,rmfa,rmfb,rmfc,rmfd,shf,shfb,shfc,shfd,shfe,shff}
or microscopic \cite{micro,micro2,micro3,micro4,micro5,micro6,micro7,micro8,vlowk,dbhf1,dbhf2,qmc} approaches for the neutron star matter EoS with hyperons.   
All these approaches agree that hyperons may appear in the inner core of neutron stars at densities of $\sim 2-3\rho_0$ as it has been said. At such densities, the nucleon chemical potential is large enough to make the conversion of nucleons into hyperons energetically favorable. This conversion relieves the Fermi pressure exerted by the baryons and makes the EoS softer, as it is illustrated in panel (a) of Fig.\ \ref{fig:EOS} for a generic model with (black solid line) and without (red dashed line) hyperons. As a consequence (see panel (b)) the mass of the star, and in particular the maximum one, is substantially reduced. In microscopic calculations  (see {\it e.g.,} Refs.\ \cite{micro,micro2,micro3,micro4,micro5,micro6,micro7,micro8,vlowk}), the reduction of the maximum mass can be even below the  "canonical'' one of $1.4-1.5M_\odot$ \cite{hulsetaylor}. This is not the case, however, of phenomenological calculations for which the maximum mass obtained is still compatible with the canonical value. In fact, most relativistic models including hyperons obtain maximum masses in the range $1.4-1.8M_\odot$ \cite{rmfa,rmfb,rmfc,rmfd}. 

Although the presence of hyperons in neutron stars seems to be energetically unavoidable, however, their strong softening of the EoS leads (mainly in microscopic models) to maximum masses not compatible with observation. 
The solution of this problem 
is not easy, and it is presently a subject of very active research, specially in view of the recent measurements of unusually high masses of the 
millisecond pulsars PSR J1903+0327 ($1.667 \pm 0.021$) \cite{freire}, PSR J1614-2230 ($1.928 \pm 0.017 M_\odot$) \cite{demorest}, and 
PSR J0348+0432 ($2.01 \pm 0.04 M_\odot$) \cite{antoniadis} 
which rule out almost all currently proposed EoS with hyperons (both microscopic and phenomenological).
To solve this problem it is necessary a mechanism that could eventually provide the additional repulsion needed to make the EoS stiffer and, therefore the maximum mass compatible with the current observational limits. Three different mechanisms that could provide such additional repulsion that have been proposed are: (i) the inclusion of a repulsive hyperon-hyperon interaction through the exchange of vector mesons \cite{Bednarek11,Weissenborn,Oertel14,Maslov},
(ii) the inclusion of repulsive hyperonic three-body forces \cite{taka,vidanatbf,yamamoto,lonardoniprl}, or (iii) the possibility of a phase transition to deconfined quark matter at densities below the hyperon threshold 
\cite{Ozel,WeissenbornSagert,Klahn2013,Bonanno,Lastowiecki2012}. In the following we briefly revise these three possible solutions. The section is finished with a short comment on the role of the $\Delta$ isobar and kaon condensation in neutron stars. 


\subsubsection{Hyperon-hyperon repulsion}
\label{subsec:yvr}

This solution has been mainly explored in the context of RMF models (see {\it e.g.,} Refs. \cite{Bednarek11,Weissenborn,Oertel14,Maslov}) and it is based on the well-known fact that,
in a meson-exchange model of nuclear forces, vector mesons generate repulsion at short distances. If the interaction of hyperons with vector mesons is repulsive enough then it could provide the required stiffness to explain the current pulsar mass observations. However, hypernuclear data indicates that, at least, the $\Lambda$N interaction is attractive \cite{hashimoto06}. Therefore, in order to be consistent with experimental data of hypernuclei, the repulsion in the hyperonic sector is included in these models only in the hyperon-hyperon interaction through the exchange of the hidden strangeness $\phi$ vector meson coupled only to the hyperons. In this way, the onset of hyperons is shifted to higher densities and neutron stars with maximum masses larger than $2M_\odot$ and a significant hyperon fraction can be successfully obtained. For further information the interested reader is referred to any of the works that have explored this solution in the last years. 


\subsubsection{Hyperonic three-body forces}
\label{subsec:yyy}

It is well known that the inclusion of three-nucleon forces  in the nuclear Hamiltonian is fundamental to reproduce properly the properties of few-nucleon systems as well as the empirical saturation point of symmetric nuclear matter in calculations based on non-relativistic many-body approaches. Therefore, it seems natural to think that three-body forces involving one or more hyperons ({\it i.e.,} NNY, 
NYY and YYY) could also play an important role in the determination of the neutron star matter EoS, and contribute to the solution of the hyperon puzzle. These forces could eventually provide, as in the case of the three-nucleon ones, the additional repulsion needed to make the EoS stiffer at high densities and, therefore, make the maximum mass of the star compatible with the recent observations. This idea was suggested even before the observation of neutron stars with $\sim 2M_\odot$ (see {\it e.g.,} Ref.\ \cite{taka}), and it has been explored by several authors in the last years \cite{vidanatbf,yamamoto,lonardoniprl}.  
However, the results of these works show that there is not yet a general consensus regarding the role of hyperonic three-body forces on the hyperon puzzle. Whereas in Refs.\ \cite{taka,yamamoto} these forces allow to obtain hyperon stars with $2M_\odot$, in Ref.\ \cite{vidanatbf} the larger maximum mass that they can support is $1.6M_\odot$, and the results of Ref.\ \cite{lonardoniprl} are not conclusive enough due to their strong dependence on the $\Lambda$NN force employed. Therefore, it seems that hyperonic three-body forces 
are not the full solution to the hyperon puzzle, although, most probably they can contribute to it in a very important way. The interested reader is referred to these works for the specific details of the calculations.

\subsubsection{Quarks in neutron stars}
\label{sec:qm_ns}

Several authors have suggested that an early phase transition from hadronic mater to deconfined quark matter at densities below the hyperon threshold could provide a solution to the hyperon puzzle. Therefore, massive neutron stars could actually be hybrid stars with a stiff quark matter core. The question that arises in this case is then whether quarks can provide the sufficient repulsion required to produce a $2M_\odot$ neutron star. To yield maximum masses larger than $2M_\odot$, quark matter should have two important and necessary features: (i) a significant overal quark repulsion resulting in a stiff EoS, and (ii) a strong attraction in a particular channel resulting in a strong color superconductivity, needed to make the deconfined quark matter phase energetically favorable over the hadronic one \cite{zdunik13}. Several models of hybrid stars with the necessary properties to generate  $2M_\odot$ neutron stars have been proposed in the recent years 
\cite{Ozel,WeissenbornSagert,Klahn2013,Bonanno,Lastowiecki2012}. Conversely, the observation of  2$M_{\odot}$ neutron stars may also helped to impose important constraints on the models of hybrid and strange stars with a quark matter core, and improve our present understanding of the hadron-quark phase transition. Here the interested reader is also referred to the original works for detail information on this possible solution.

\subsubsection{$\Delta$ isobar and kaon condensation in neutron stars}
\label{subsec:delta}

An alternative way to circumvent the hyperon puzzle is to invoke the appearance of other hadronic degrees of freedom such as for instance the $\Delta$ isobar or meson condensates that push the onset of hyperons to higher densities. 

Usually, the $\Delta$ isobar is neglected in neutron stars since its threshold density was found to be higher than the typical densities prevalent in the neutron star core. However, this possibility has been recently reviewed by Drago {\it et al.,} in Ref.\ \cite{Drago14}. The authors of this work have shown that the onset of the $\Delta$ depends crucially on the density-dependence of the derivative parameter of the nuclear symmetry energy, $L=3\rho_0(\partial E_{sym}(\rho)/\partial \rho)_{\rho_0}$. By using a state-of-the-art EoS and recent experimental constraints of $L$, these authors showed that the $\Delta$ isobar could actually appear before the hyperons in the neutron star interior. However, they found that, as soon as the $\Delta$ is present the EoS, as in the case of hyperons, becomes considerably softer and, consequently, the maximum mass is reduced to values below the current observational limit also in this case, giving rise to what has been 
recently called the $\Delta$ puzzle.

The possible existence of a Bose--Einstein condensate of negative kaons in the inner core of neutron stars has also been also been extensively considered in the literature (see {\it e.g.,} \cite{kaon1,kaon1b,kaon2,kaon3,kaon4,kaon5} and references therein). As the density of stellar matter increases, the $K^-$ chemical potential, $\mu_{K^-}$, is lowered by the attractive vector meson field originating from dense nucleonic mater. When $\mu_{K^-}$ becomes smaller than the electron chemical potential $\mu_{e}$ the process $e^-\rightarrow K^-+\nu_e$ becomes energetically possible. The critical density for this process was calculated to be in the range $2.5-5\rho_0$ \cite{kaon3,kaon4}. However, as in the case of the $\Delta$, the appeareance of the kaon condensation induces also a strong softening of the EoS and the consequently leads to a reduction of the maximum mass to values also below the current observational limits. The interested reader to the original works on this subject \cite{kaon1,kaon1b,kaon2,kaon3,kaon4,kaon5} for 
a comprehensive description of the implications of kaon condensation on the structure and evolution of neutron stars.


\subsection{Hyperon stars at birth and neutron star cooling}
\label{sec:cool}

As it is said at the beginning of this section, neutron stars are formed in Type-II, Ib or Ic  supernova explosions. Properties of newly born neutron stars are affected by thermal effects and neutrino trapping. These two effects have a strong influence on the overall stiffness of the EoS and the composition of the star. In particular (see {\it e.g.,} \cite{keil,keil2,keil3,vidanaAA,burgio}) matter becomes more proton rich, the number of muons is significantly reduced, and the onset of hyperons is shifted to higher densities. In addition, the number of strange particles 
is on average smaller, and the EoS is stiffer in comparison with the cold and neutrino-free case.

\begin{figure*}[t]
\begin{center}
\resizebox{0.90\textwidth}{!}
{
\includegraphics[clip=true]{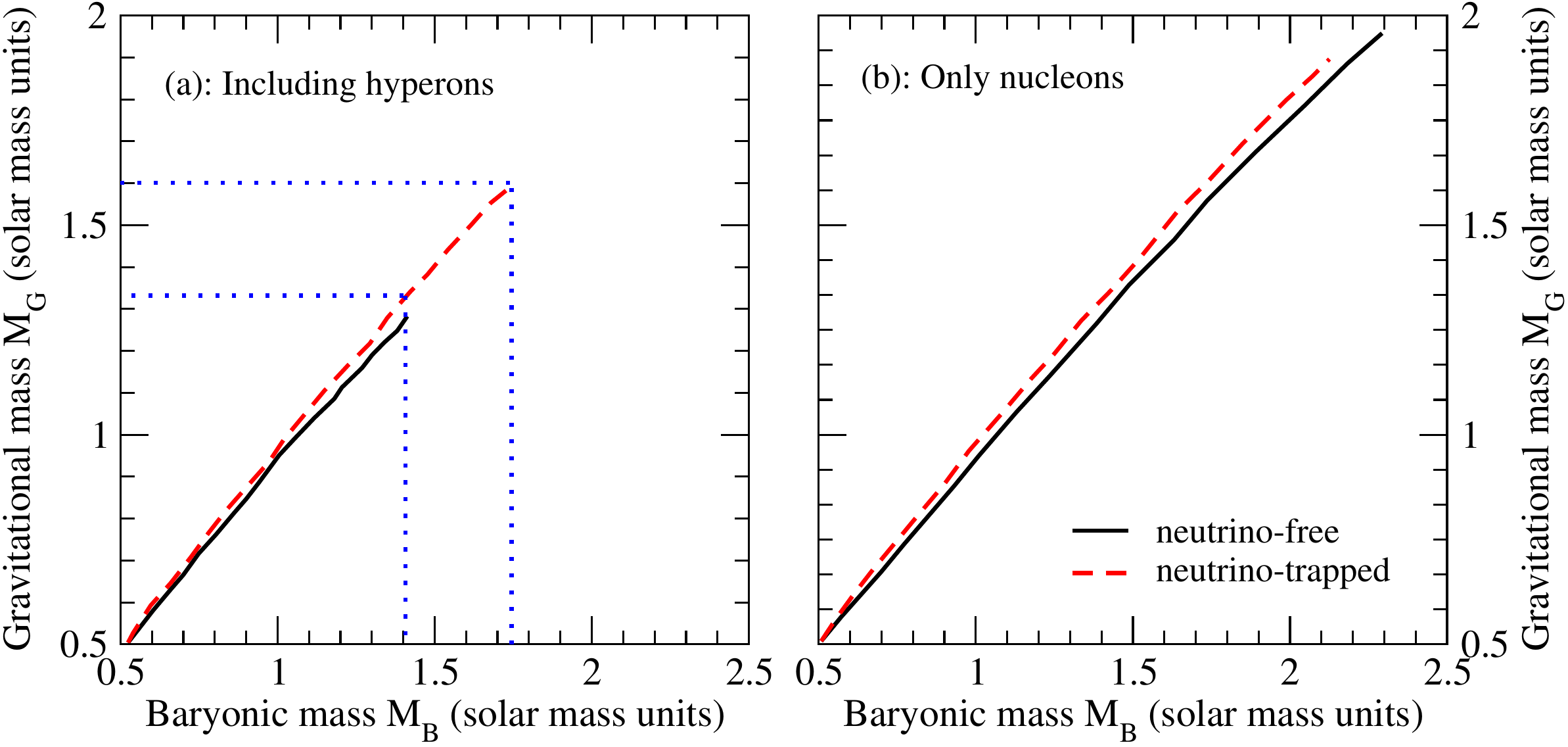}
}
\caption{Gravitational mass as a function of the baryonic mass for neutrino-free (solid lines) and neutrino-trapped (dashed lines) matter. Panel (a) shows the results for matter containing nucleons and hyperons, whereas the results for pure nucleonic mater are shown in panel (b). Dotted horizontal and vertical lines show the window of metastability in the gravitational and baryonic masses. Figure adapted from Ref.\ \cite{vidanaAA}.} 
\label{fig:MGMB}       
\end{center}
\end{figure*}

A very important implication of neutrino trapping in dense matter is the possibility of having metastable  neutron stars
and a delayed formation of a "low-mass'' ($M=1-2M_\odot$) black hole. This is illustrated in Fig.\ \ref{fig:MGMB} for the case 
of the BHF calculation of Ref.\ \cite{vidanaAA}. The figure shows the gravitational mass $M_G$ of the star as a function of its baryonic mass $M_B$. If 
hyperons are present (panel (a)), then deleptonization lowers the range of gravitational masses that can be supported by the EoS from about $1.59 M_\odot$ 
to about $1.28 M_\odot$ (see dotted horizontal lines in the figure). Since most of the matter accretion on the forming neutron star happens in a very early stages
after birth ($t<1$ s), with a good approximation, the neutron star baryonic mass stays constant during the evolution from the initial proto-neutron star configuration
to the final neutrino-free one. Then, for this particular model, proto-neutron stars which at birth have a gravitational mass between $1.28-1.59 M_\odot$ (a baryonic 
mass between $1.40-1.72 M_\odot$) will be stabilized by neutrino trapping effects long enough to carry out nucleosynthesis accompanying a Type-II supernova 
explosion. 
After neutrinos leave the star, the EoS is softened and it cannot support anymore the star against its own gravity. The newborn star collapses then to a 
black hole \cite{keil,keil2,keil3}. On the other hand, if only nucleons are considered to be the relevant baryonic degrees of freedom 
(panel (b)), no metastability occurs and a black hole is unlikely to be formed during the deleptonization since the gravitational mass increases during 
this stage which happens at (almost) constant baryonic mass. If a black hole were to form from a star with only nucleons, it is much more likely to form during 
the post-bounce accretion stage.


The cooling of the newly born hot neutron stars is driven first by the neutrino emission from the interior,
and then by the emission of photons at the surface. Neutrino emission processes can be divided into slow and fast processes depending on
whether one or two baryons participate. The simplest possible neutrino emission process is the so-called direct Urca process:
\begin{equation}
n \rightarrow p+l+\bar \nu_l \ , \,\,\, p+l \rightarrow n +\nu_l \ .
\end{equation}
This is a fast mechanism which however, due to momentum conservation, it 
is only possible when the proton fraction exceeds a critical value $x_{DURCA} \sim 11\%$ to $15 \%$ \cite{lattimer}. Other neutrino processes
which lead to medium or slow cooling scenarios, but that are operative at any density and proton fraction, are the so-called modified Urca processes:
\begin{equation}
N+ n \rightarrow N+ p+l+\bar \nu_l \ , \,\,\, N+p+l \rightarrow N+n +\nu_l \ , 
\end{equation}
the bremsstrahlung: 
\begin{equation}
N+N \rightarrow N+N + \nu +\bar \nu \ ,
\end{equation}
or
the Cooper pair formation: 
\begin{equation}
n+n\rightarrow [nn]+\nu+\bar \nu \ , \,\,\, p+p\rightarrow [pp]+\nu+\bar \nu, 
\end{equation}
this last operating only when the
temperature of the star drops below the critical temperature for neutron superfluidity or proton superconductivity. If hyperons are present in the neutron star interior new 
neutrino emission processes, like {\it e.g.,} 
\begin{equation}
Y\rightarrow B+l+\bar\nu_l \ , 
\end{equation}
may occur providing additional fast cooling mechanisms.  Such additional rapid cooling mechanisms, however, can lead to surface temperatures much lower than 
that observed, unless they are suppressed by hyperon pairing gaps. Therefore, the study of hyperon superfluidity becomes of particular interest since it 
could play a key role in the thermal history of
neutron stars. Nevertheless, whereas the presence of superfluid neutrons in the inner crust of neutron stars, and superfluid neutrons together with
superconducting protons in their quantum fluid interior is well established and has been the subject of many studies, a quantitative estimation of the
hyperon pairing has not received so much attention, and just few calculations exists in the literature \cite{super,super2,super3,super4,super5,super6,super7}.
 

\subsection{Hyperons and the r-mode instability of neutron stars}
\label{sec:rmode}

It is well known that the upper limit on the rotational frequency of a neutron star is set by its
Kepler frequency $\Omega_{Kepler}$, above which matter is ejected from the star's equator
\cite{lindblom86,lindblom86b}. However, a neutron star may be unstable against some perturbations which
prevent it from reaching rotational frequencies as high as $\Omega_{Kepler}$, setting, therefore,
a more stringent limit on its rotation \cite{lindblom85}. Many different instabilities can operate in
a neutron star. Among them, the so-called r-mode instability \cite{anderson,andersonb}, a toroidal mode of
oscillation whose restoring force is the Coriolis force, is particularly interesting. This oscillation mode 
leads to the emission of gravitational waves in hot and rapidly rotating neutron stars though the 
Chandrasekhar--Friedman--Schutz mechanism \cite{cfs,cfsb,cfsc,cfsd}. Gravitational
radiation makes an r-mode grow, whereas viscosity stabilizes it. Therefore, an r-mode is unstable
if the gravitational radiation driving time is shorter than the damping time due to viscous processes.
In this case, a rapidly rotating neutron star could transfer a significant fraction of its rotational energy
and angular momentum to the emitted gravitational waves. The detection of these gravitational waves could
provide invaluable information on the internal structure of the star and constraints on the EoS.

Bulk ($\xi$) and shear ($\eta$) viscosities are usually considered the main dissipation mechanism of r- and other pulsation
modes in neutron stars. Bulk viscosity is the dominant one at high temperatures ($T> 10^9$ K) and, therefore,
it is important for hot young neutron stars. It is produced when the pulsation modes induce variations in pressure
and density that drive the star away from $\beta$-equilibrium. As a result, energy is dissipated as the weak
interaction tries to reestablish the equilibrium. In the absence of hyperons or other exotic components, the bulk
viscority of neutron star matter is mainly determined by the reactions of direct  and modified  Urca processes. 
However, has soon as hyperons appear new mechanisms such as weak non-leptonic hyperon reactions: 
\begin{equation}
N+N\leftrightarrow N+Y \ , \,\,\, N+Y \leftrightarrow Y+Y \ , 
\end{equation}
direct  and modified hyperonic Urca: 
\begin{equation}
Y \rightarrow B+l+\bar \nu_l \ , \,\,\, B+l \rightarrow Y +\nu_l \ ,
\end{equation}
\begin{equation}
B'+Y \rightarrow B'+B+l+\bar \nu_l \ , \,\,\, B'+B+l \rightarrow B'+ Y +\nu_l \ ,
\end{equation}
or strong interactions: 
\begin{equation}
Y+Y \leftrightarrow N+Y \ , \,\,\, N+\Xi \leftrightarrow Y+Y\ , \,\,\, Y+Y \leftrightarrow Y+Y
\end{equation}
contribute to the bulk viscosity and dominate it for $\rho \geq 2-3 \rho_0 $. Several works have been devoted to the study of the hyperon bulk
viscosity \cite{ybv,ybv2,ybv3,ybv4,ybv5,ybv6,ybv7,ybv8,ybv9,ybv10,ybv11,ybv12,ybv13,ybv14,ybv15}. The interested reader is referred to these works for detailed studies on this topic. 

The time dependence of an r-mode oscillation is given by $e^{i\omega t-t/\tau(\Omega,T)}$, where $\omega$ is the frequency of the mode, and $\tau(\Omega, T)$ is an 
overall time scale of the mode which describes both its exponential growth due to gravitational wave emission as well as its decay due to viscous damping.
\begin{figure}[t]
\begin{center}
\resizebox{0.90\textwidth}{!}
{
\includegraphics[clip=true]{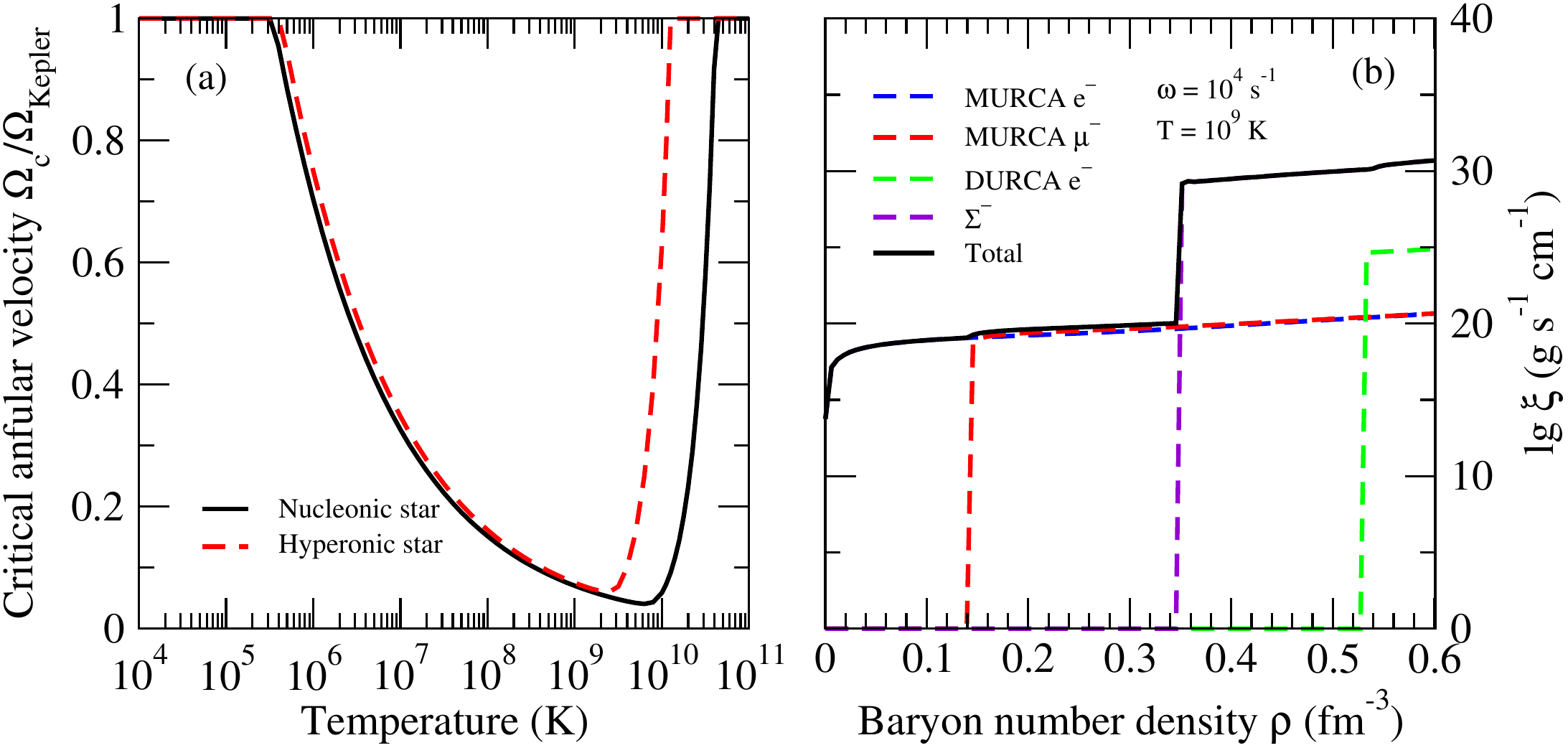}
}
\caption{
Panel (a): r-mode instability region for a pure nucleonic and a hyperonic star with $1.27 M_\odot$. The frequency of the mode is taken as
$\omega=10^4$ s$^{-1}$.
Panel (b):
Bulk viscosity as a function of the density for $T=10^9$ K and $\omega=10^4$ s$^{-1}$. Contributions 
direct and modified nucleonic Urca processes as well as from the weak non-leptonic process  $n+n\leftrightarrow p+\Sigma^-$ are included. Figure adapted from Ref.\ \cite{albertus}.} 
\label{f:fig3}       
\end{center}
\end{figure}
It can be written as 
\begin{equation}
\frac{1}{\tau(\Omega, T)}=-\frac{1}{\tau_{GW}(\Omega)}+\frac{1}{\tau_{\xi}(\Omega,T)}+\frac{1}{\tau_{\eta}(\Omega,T)} \ . 
\end{equation}
If $\tau_{GW}$ is shorter than both $\tau_\xi$ and $\tau_\eta$ the mode will 
exponentially grow, whereas in the opposite case it will be quickly damped away. For each star at a given temperature T one can define a critical angular
velocity $\Omega_c$ as the smallest root of the equation
\begin{equation}
\frac{1}{\tau(\Omega_c, T)}=0 \ . 
\end{equation}
This equation defines the boundary of the so-called r-mode instability region.
A star will be stable against the r-mode instability if its angular velocity is smaller than its corresponding $\Omega_c$. On the contrary, a star with 
$\Omega > \Omega_c$ will develope an instability that will cause a rapid loss of angular momentum through gravitational radiation until its angular velocity
falls below the critical value. On  panel (a) of Fig.\ \ref{f:fig3} it is presented,  as example, the r-mode instability region for a pure nucleonic (black solid line) and a hyperonic 
(red dashed line) star with $1.27 M_\odot$ \cite{albertus}. 
The contributions to the bulk viscosity from 
direct and modified nucleonic Urca processes as well as from the weak non-leptonic process  $n+n\leftrightarrow p+\Sigma^-$ included in the calculation are shown in the panel (b) of the figure.
Clearly the r-mode instability is smaller for the hyperonic star. The reason being simply  the increase of the bulk viscosity 
due to the presence of hyperons which makes the damping of the mode more efficient.


\section{Summary and Conclusions}
\label{sec:conclusions}

In this review article we have discussed several topics of hypernuclear physics. After a short introduction to the field  we have discussed in the first part of this work different production mechanism of single- and double-$\Lambda$ hypernuclei, as well as several aspects of $\gamma$-ray hypernuclear spectroscopy and weak decay modes of hypernuclei. Then, we have reviewed several approaches to build the hyperon-nucleon interaction. In particular, we have discussed models for the hyperon-nucleon interaction based on meson-exchange theory, chiral effective field theory and the recent 
$V_{low\,\, k}$ approach and lattice QCD developments. Finally, we have discussed the main effects of hyperons on the properties of neutron stars with an emphasis on the so-called "hyperon puzzle'', {\it i.e.,} the problem of the strong softening of the EoS of dense matter due to the appearance of hyperons which leads to maximum masses of compact stars that are not compatible with the recent observations of $\sim 2 M_\odot$ millisecond pulsars. We have discussed three different solutions proposed to tackle this problem: (i) more repulsion in hyperon-hyperon interactions within the density functional theories
of hypernuclear matter in the vector and/or scalar mesons exchange channels; (ii) repulsive hyperonic three-body forces in the ab initio microscopic calculations, and (iii) a phase transition to deconfined quark matter at densities below the hyperon threshold.  The role of $\Delta$ isobar on the possible solution of this problem as also been revised. We have also presented a discussion of how the presence of hyperons will affect the cooling of neutron stars and the r-mode instability window through modifications of the microscopic input of the  weak interaction rates and transport coefficients of dense matter.


\section*{Acknowledgements}

The author is grateful to Assum Parre\~no for her useful comments on the hypernuclear weak decay and the lattice QCD sections, and to D. Unkel for the interesting discussions they had during the development of this work. This work is supported by "PHAROS: The multi-messenger physics and astrophysics of neutron stars'', COST Action CA16214.


\end{document}